\documentclass[fleqn,usenatbib]{mnras}

\usepackage[T1]{fontenc}

\DeclareRobustCommand{\VAN}[3]{#2}
\let\VANthebibliography\thebibliography
\def\thebibliography{\DeclareRobustCommand{\VAN}[3]{##3}\VANthebibliography}

\usepackage{graphicx}	
\usepackage{amsmath}	
\usepackage{amssymb}	
\usepackage{booktabs}
\usepackage{subfigure}

\def\HI{H\textsc{i}}
\newcommand{\msol}{\text{M}$_\odot$}
\newcommand{\U}{U$_\text{336}$}
\newcommand{\B}{B$_\text{438}$}
\newcommand{\BA}{B$_\text{435}$}
\newcommand{\V}{V$_\text{606}$}
\newcommand{\I}{I$_\text{814}$}
\newcommand{\Tcr}{$T_\text{cr}$}
\newcommand{\rh}{$r_\text{h}$}

\title[Star Clusters in Tidal Debris]{Star Clusters in Tidal Debris}

\author[Michael Rodruck et al.]{
	Michael Rodruck,$^{1,2}$\thanks{E-mail: mrodruck@gmail.com}
	Jane Charlton,$^{1}$
	Sanchayeeta Borthakur,$^{3}$ 
	Aparna Chitre,$^{4}$  \newauthor
	Patrick R. Durrell,$^{5}$
	Debra Elmegreen,$^{6}$  
	Jayanne English,$^{7}$  
	Sarah C. Gallagher,$^{8}$  
	Caryl Gronwall,$^{1,9}$\newauthor
	Karen Knierman,$^{3}$
	Iraklis Konstantopoulos,$^{10}$  
	Yuexing Li,$^{1}$ 
	Moupiya Maji,$^{11}$
	Brendan Mullan,$^{12}$\newauthor
	Gelys Trancho$^{13}$ and
	William Vacca$^{14}$ 
	\\
	$^{1}$Department of Astronomy and Astrophysics, The Pennsylvania State University, University Park, PA 16802, USA\\
	$^{2}$Department of Physics, Engineering, and Astrophysics, Randolph-Macon College, Ashland, VA 23005, USA\\
	$^{3}$School of Earth and Space Exploration, Arizona State University, 550 E. Tyler Mall, Room PSF-686 (PO Box 871404), Tempe, AZ\\ 85287-1404, USA\\
	$^{4}$Space Telescope Science Institute, 3700 San Martin Drive, Baltimore, MD 21218, USA\\
	$^{5}$Department of Physics, Astronomy, Geology and Environmental Sciences, Youngstown State University, Youngstown, OH 44555, USA\\
	$^{6}$Department of Physics and Astronomy, Vassar College, Poughkeepsie, NY 12604, USA\\
	$^{7}$Department of Physics and Astronomy, University of Manitoba, Winnipeg, MB R3T 2N2, Canada\\
	$^{8}$Department of Physics and Astronomy and Institute for Earth and Space Exploration, University of Western Ontario,\\ London, ON N6A 3K7, Canada\\
	$^{9}$Institute for Gravitation and the Cosmos, The Pennsylvania State University, University Park, PA 16802, USA\\
	$^{10}$Independent scholar, Wellington, New Zealand\\
	$^{11}$Inter-University Centre for Astronomy and Astrophysics (IUCAA), Post Bag 4, Ganeshkhind, Pune 411 007, India\\
	$^{12}$Department of Natural Sciences and Engineering Technology , Point Park University 201 Wood St, Pittsburgh, PA 15222, USA\\
	$^{13}$Thirty Meter Telescope International Observatory, 100 W Walnut St, \#300, Pasadena, CA 91124, USA\\
	$^{14}$SOFIA-USRA, NASA Ames Research Center, MS N232-12, Moffett Field, CA 94035-1000, USA
}

\date{Accepted XXX. Received YYY; in original form ZZZ}

\pubyear{2021}

\begin{document}
	\label{firstpage}
	\pagerange{\pageref{firstpage}--\pageref{lastpage}}
	\maketitle
	
	\begin{abstract}
		We present results of a \textit{Hubble Space Telescope} (\textit{HST}) \textit{UBVI}-band study of star clusters in tidal tails, using new WFC3 and ACS
		imaging to complement existing WFPC2 data. We survey 12 tidal tails
		across seven merging systems, deriving ages and masses for 425 star cluster candidates (SCCs). 
		The stacked mass distribution across all systems follows a power law of the form $dN/dM \propto  M^{\beta}$, with $\beta = -2.02 \pm 0.15$,
		consistent with what is seen in other star forming environments. \textit{GALEX} and \textit{Swift} UV imaging provide star formation
		rates (SFRs) for our tidal tails, which when compared with ages and masses of our SCCs, allows for a determination of the
		cluster formation efficiency (CFE). We find the CFE increases with increasing SFR surface density, matching the theoretical model.
		We confirm this fit down at SFR densities lower than previously measured (log $\Sigma_\text{SFR} \: (\text{M}_\odot \: 
		\text{yr}^{-1} \: \text{kpc}^{-2})
		\approx -4.2$), as related to the CFE. We determine the half-light radii for a refined sample of 57 SCCs with our \textit{HST} WFC3 and ACS imaging,
		and calculate their dynamical age, finding the majority of them to be gravitationally bound.
		We also provide evidence of only low-mass ($< 10^4$ \msol) cluster formation in our nearest galaxy, NGC 1487, consistent with the theory that this
		system is a dwarf merger. 
		
	\end{abstract}
	
	
	
	
	\section{Introduction}

	The gravitational origin of tidal tails was first realized with the pivotal work by \cite{toomre_72}, who showed that galactic tidal tails and warped
	disks could be simulated as gravitational encounters with another galaxy. In a very forward thinking
	section of their work, they suggested that galaxy mergers may be able to produce large amounts of star formation. This prediction was later
	borne out with the discovery of luminous infrared galaxies (LIRGs), made possible with the launch of the IR telescope \textit{IRAS}. \textit{IRAS}
	found that many galaxies with high ($> 10^{11} L_\odot$) IR luminosites have disturbed morphologies,
	indicative of past merging events \citep{sanders_88}. The IR emission of these LIRGs suggested star formation rates (SFRs) on the order of
	100 \msol / yr \citep{schweizer_87_2}.
	
	Galaxy mergers can produce collisions between clouds of gas, which can provoke star formation. The high pressures generated in these
	collisions will lead to enhanced star formation efficiency \citep{jog_92} and the formation of massive star clusters \citep{zubovas_14,maji_17}. Additionally,
	cloud collisions may provide external pressure to young clusters, keeping their contents gravitationally bound, preventing their destruction
	\citep{elm_08}. The advent of high-resolution imaging, possible with \textit{HST}, has found evidence of such massive clusters within the interiors
	of merging galaxies in the form of bright, blue, compact objects (e.g. \citealp{zepf_99,whitmore_95,whitmore_93}). These objects, labeled as young
	massive clusters (YMCs), show properties similar to what is expected for young globular clusters, such as mass and radius. Therefore, by studying 
	YMCs, we may be able to study the formation of today's globular clusters.
	
	While the interiors of mergers have been well-studied in the past, few surveys have looked at the tidal debris associated with mergers.
	However, simulations are now showing that star formation can occur in the extended regions of a merger. Simulations with
	explicit stellar feedback indicate that $\sim20 - 50\%$ of a merger's SFR can occur in extended debris \citep{hopkins_13}. Observations 
	of the Tadpole galaxy confirm this prediction, as $\sim30\%$ of the system's star formation is occurring in tidal tail star clusters \citep{jarrett_06}. Tidal
	tails also offer a relatively clean and uncluttered environment as compared to the interiors of galaxies. Their sparse environments mean clusters will 
	not be subject to shocks and tides found in the dense nuclear region, and may avoid disruption,
	surviving to the present day \citep{renaud_18}.
	
	Despite the differences in location and interaction, several common properties seem to exist between studies of YMCs found in merging and quiescent 
	galaxies. The slopes of the mass
	and luminosity functions has been measured to be between $-1.8$ and $-2.2$ for both types of galaxies. The cluster formation efficiency, a measure of the 
	percentage of the SFR occurring within star clusters, tracks the local SFR density in both environments. Additionally, cluster radii are similar, with half-light radii of $\sim 0.5 - 10$ pc \citep{zwart_10}.
	Such similarities suggests common physics behind star cluster formation.
	
	This paper builds on previous work by \cite{mullan_11} and \cite{knierman_03}, who studied clusters in tidal tails using the 
	WFPC2 camera on \textit{HST}. Both works found statistically significant populations of clusters in a variety of tails, with colours suggesting
	masses comparable to the YMCs found in the interiors of mergers. Merging systems with young interaction ages and bright
	tails produced the most star clusters. However, analysis was limited by the lack of multiband photometry.	
	Our new observations of 12 tails in seven interacting systems add F336W and F438W WFC3 and F435W ACS imaging to existing WFPC2 F606W and F814W observations,
	to allow for age and mass determination.
	

	We will begin in Section \ref{sec:2_2} by describing our imaging datasets and our analysis methods. In Section \ref{sec:3_2} 
	we show our results. In Section
	\ref{sec:4_2} we discuss our findings, and conclude our chapter with the main points of our research in
	Section \ref{sec:5_2}. We add notes for individual tails in Section \ref{sec:6_2}.
	
	%

	\section{Data Analysis} \label{sec:2_2}
	Our analysis consists of two parts: \textit{HST} imaging to identify
	tidal tail star clusters and determine their ages and masses, and UV imaging
	with \textit{GALEX} and \textit{Swift} to derive the local star formation rates.
	Both efforts are described below.
	
	\subsection{\textit{HST}}
\begin{figure*}
		\centering
		\includegraphics[width=0.8\linewidth]{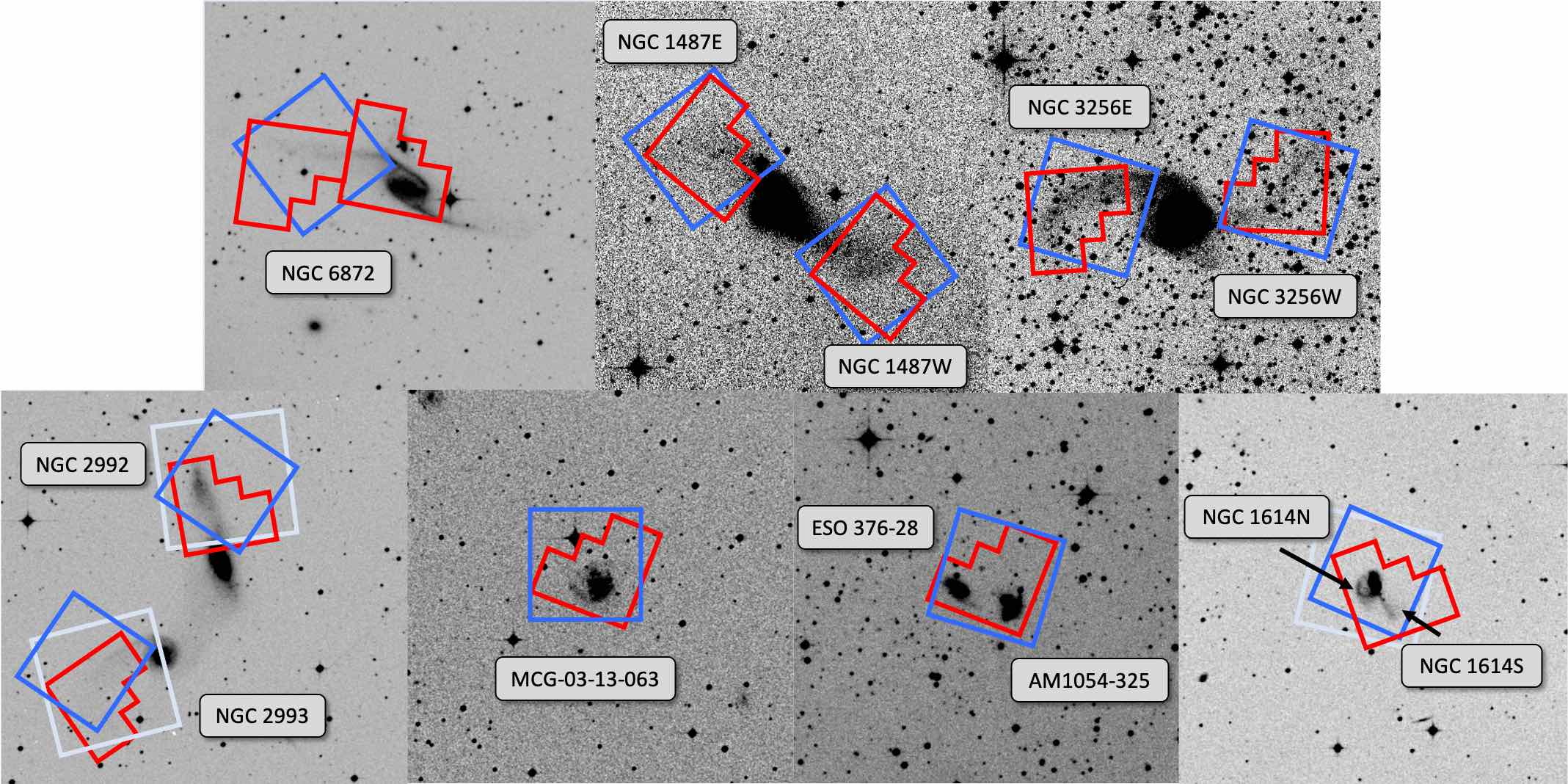}
		\caption{Snapshots of our sample from the Digitized Sky Survey (DSS). Red outlines indicate WFPC2 pointings from \protect\cite{mullan_11} and \protect\cite{knierman_03}. 
			Blue and cyan squares show the WFC3 and ACS footprints, respectively.}
		\label{fig:sample}
	\end{figure*}

	Our optical and near-IR observations consist of WFPC2, ACS, and WFC3 imaging from \textit{HST} across multiple cycles; the full sample is shown
	in Figure \ref{fig:sample}. Properties of our sample are highlighted in Table \ref{table:data},
	with systems ordered according to their interaction age. We sample not only major disk mergers, but minor 
	and dwarf mergers as well.

\begin{table*}
	\caption{System information. Interaction age gives the most recent interaction, which produced the visible tidal features.}
	\begin{tabular}{llllll}
		\textbf{System} & \textbf{Interaction Age} & \textbf{Distance} & \textbf{Merger Type} & \textbf{Tidal Features} & \textbf{Tidal Features $\mu_V$} \\
		& (Myr)                    & (Mpc)            &                      &                        & (mag arcsec$^{-1}$) \\ \hline \hline
		NGC 1614N/S      & 50                       & 65.6             & Major                & Tidal tails            & 22.27/23.00          \\
		AM1054-325/ESO 376-28 & 85                  & 52.9             & Major                & Tidal tails and tidal dwarf & 22.65/22.76 \\
		NGC 2992/3       & 100                      & 36.6             & Major                & Tidal tails and tidal bridge & 23.47/24.78 \\
		MCG-03-13-063    & 100                      & 46.2             & Minor                & Extended spiral arm    & 23.91                \\
		NGC 6872         & 150                      & 62.6             & Minor                & Tidal tails            & 24.06                \\
		NGC 3256E/W      & 400                      & 42.8             & Major                & Tidal tails            & 24.04/23.75          \\
		NGC 1487E/W      & 500                      & 10.8             & Dwarf                & Tidal tails            & 24.02/24.57          \\ \hline
	\end{tabular}
	
	\label{table:data}
\end{table*}

	WFPC2 F555W and
	F814W data were taken in Cycle 7 (GO 7466) \cite{knierman_03} for NGC 3256. The remaining samples of AM1054-325, MCG-03-068-13, NGC 1487, NGC 2992, NGC 2993, NGC 6872, and NGC 1614 were observed with WFPC2 in Cycle 16 (GO 11134), in
	F606W and F814W \cite{mullan_11}.
	Galaxies from \cite{mullan_11} represent an extension of the original sample in \cite{knierman_03}, designed to sample a variety of ages, mass ratios, and optical
	properties. Our survey adds WFC3 F336W and WFC3 F438W/ACS F435W imaging to galaxies lying in the Southern hemisphere, which have also been
	observed with the Gemini-South GMOS detector, and will be the subject of a forthcoming paper on the diffuse light in the tidal debris. Archival ACS F438W and WFC3 F336W data for NGC 1614N/S
	are used from Cycle 14 (GO 10592) and Cycle 23 (GO 14066).
	A full description of our \textit{HST} observations is in Table \ref{table:exposures_2}. We refer to our F606W, F814W, F336W, F438W, and F435W observations as \V, \I, \U,
	\B, and \BA, respectively.

	WFPC2 data were reduced in IRAF and corrected for
	degraded charge transfer effects (CTE) using \cite{dolphin_00}. We refer the reader to the respective papers
	\citep{mullan_11,knierman_03} for a more in depth discussion on WFPC2 data reduction.
	WFPC2 magnitudes were taken from \cite{mullan_11} and \cite{knierman_03}. Objects in \cite{knierman_03} were transformed from the WFPC2 Vegamag
	photometric system to the Johnson-Cousins system using transformations found in \cite{holtzman_95}, going from F555W and F814W to \textit{V}
	and \textit{I}. Rather than converting back to WFPC2 magnitudes, we keep them in their transformed system. Data from \cite{mullan_11} were provided in WFPC2
	Vegamag magnitudes (\V\ and \I).
	
	ACS and WFC3 images were downloaded from the \textit{HST} archive, which have been processed through the standard ACS and WFC3 pipelines. 
	ACS images have a scale of 0.05'' per pixel, while WFC3 observations have a scale of 0.04'' per pixel. This is compared to the WFPC2 WF scale
	of 0.1'' per pixel; objects on the WFPC2 PC were not analyzed due to its larger readnoise.
	
	We perform aperture photometry on all our objects using the DAOPHOT.PHOT \citep{stetson_87} task in IRAF. We choose to match the aperture settings
	in \cite{mullan_11}, which uses radii of 2, 5, and 8 pixels for object, inner background, and outer background annuli, respectively. When
	translated to the WFC3 scale, this results in radii of 5, 12.5, and 20 pixels. For ACS, we use radii of 4, 10, and 16 pixels.
	In this manner, we are assured that we are measuring our clusters to the same physical extent, camera to camera. 
	Aperture corrections to bright, isolated stars were performed out to a radius of 10 pixels for WFC3 and ACS imaging. Due
	to the lack of adequate stars in the NGC 1487 field, we take the average aperture corrections from all other fields and apply them to NGC 1487W and NGC 1487E.
	10 pixel zeropoints of
	23.392, 24.895, and 25.762 were used for \U, \B, and \BA, respectively.

	

		\begin{table*}
			\caption{\textit{HST} observations. \hspace{\linewidth} }
			\begin{center}
		\begin{tabular}{llllll}
			System & Filter & Exposure time (s) & Date & Program ID & Camera \\ \hline \hline
			NGC 1614N/S & F336W & 6510 & 2015 Dec 12 & 14066 & WFC3 \\
			& F438W & 1260 & 2006 Aug 14 & 10592 & ACS \\
			& F606W & 1900 & 2007 Nov 15 & 11134 & WFPC2 \\
			& F814W & 1900 & 2007 Nov 15 & 11134 & WFPC2 \\ 
			AM 1054-325/ESO 376-28 & F336W & 3440 & 2017 Nov 6 & 14937 & WFC3 \\
			& F435W & 1520 & 2017 Nov 7 & 14937 & WFC3 \\
			& F606W & 1900 & 2008 Feb 24 & 11134 & WFPC2 \\
			& F814W & 1900 & 2008 Feb 24 & 11134 & WFPC2 \\
			NGC 2992 & F336W & 2230 & 2018 Nov 27 & 15083 & WFC3 \\
			& F438W & 2100 & 2018 Apr 19 & 15083 & ACS \\
			& F606W & 1000 & 2007 Dec 28 & 11134 & WFPC2 \\
			& F814W & 900 & 2007 Dec 28 & 11134 & WFPC2 \\
			NGC 2993 & F336W & 2230 & 2018 Apr 19 & 15083 & WFC3 \\
			& F438W & 2100 & 2018 Nov 27 & 15083 & ACS \\
			& F606W & 1000 & 2007 Dec 3 & 11134 & WFPC2 \\
			& F814W & 900 & 2007 Dec 3 & 11134 & WFPC2 \\
			MCG-03-13-063 & F336W & 3420 & 2017 Sep 27 & 14937 & WFC3 \\
			& F435W & 1520 & 2017 Sep 28 & 14937 & WFC3 \\
			& F606W & 1000 & 2007 Nov 24 & 11134 & WFPC2 \\
			& F814W & 900 & 2007 Nov 24 & 11134 & WFPC2 \\
			NGC 6872$^a$ & F336W & 3920 & 2018 Jul 24 & 15083 & WFC3 \\
			& F435W & 1760 & 2018 Jul 24 & 15083 & WFC3 \\
			& F606W & 2100/1900 & 2008 Feb 23/2008 May 16 & 11134 & WFPC2 \\
			& F814W & 2100/1900 & 2008 Feb 23/2008 May 16 & 11134 & WFPC2 \\
			NGC 3256E & F336W & 3520 & 2018 Jun 15 & 15083 & WFC3 \\
			& F435W & 1600 & 2018 Jun 15 & 15083 & WFC3 \\
			& F555W & 1000 & 1999 Oct 11 & 7466 & WFPC2 \\
			& F814W & 1000 & 1999 Oct 11 & 7466 & WFPC2 \\
			NGC 3256W & F336W & 3520 & 2018 Jan 15 & 15083 & WFC3 \\
			& F435W & 1600 & 2018 Jan 15 & 15083 & WFC3 \\
			& F555W & 1000 & 1999 Mar 24 & 7466 & WFPC2 \\
			& F814W & 1000 & 1999 Mar 24 & 7466 & WFPC2 \\
			NGC 1487E & F336W & 3520 & 2017 Nov 25 & 15083 & WFC3 \\
			& F435W & 1600 & 2017 Nov 25 & 15083 & WFC3 \\
			& F606W & 1000 & 2008 Aug 9 & 11134 & WFPC2 \\
			& F814W & 900 & 2008 Aug 9 & 11134 & WFPC2 \\
			NGC 1487W & F336W & 3520 & 2019 Mar 24 & 15083 & WFC3 \\
			& F435W & 1600 & 2019 Mar 24 & 15083 & WFC3 \\
			& F606W & 1000 & 2008 Aug 31 & 11134 & WFPC2 \\
			& F814W & 900 & 2008 Aug 31 & 11134 & WFPC2 \\ \hline
		\end{tabular}
			\end{center}
		\raggedright
		$^a$NGC 6872 was measured in two pointings with WFPC2.
		\label{table:exposures_2}
	\end{table*}

	\subsection{\textit{GALEX}}
	The \textit{Galaxy Evolution Explorer} (\textit{GALEX}) operated from 2003 to 2012 as an orbiting UV telescope. Its 50 cm
	diameter mirror imaged a 1.25$^{\circ}$ FOV with two separate microchannel-plate detectors, designated as the near-ultraviolet (NUV) and
	far-ultraviolet (FUV) channels. The telescope has a pixel scale of 1.5'' per pixel.
	GALEX conducted several surveys, including the All-Sky Imaging survey (AIS), Medium Imaging Survey (MIS), Deep Imaging Survey (DIS),
	and Nearby Galaxy Survey (NGS), to varying depths. Our targets are drawn from AIS (AM1054-325, ESO 376-28, MGC-03-13-063, NGC 1487E/W)
	and NGS (NGC 6872, NGC 2992, NGC 2993). Our \textit{GALEX} observations are listed in Table \ref{UV}, along with our
	\textit{Swift} observations (described below).
	
	We use images taken in the FUV ($\lambda_{\text{eff}} = 1538.6 \AA$) obtained from the MAST archive. These have been pipeline processed.
	While background subtracted images are available via MAST, we do not use these because the background subtraction can be inaccurate for extended objects.
	Photometry was performed on the entire flux enclosed within the tail boundary, as described in Section \ref{sec:def}. The sky background was determined 
	by sampling five nearby regions of 30 x 30 pixels and taking the mean value. For NGC 2992, a bright, blue star visible in the FUV image was manually masked.
	
	We corrected for Galactic extinction using coefficients from \cite{yuan_13} for \textit{GALEX} \textit{FUV}.
	
	\begin{table*}
				\caption{\textit{GALEX} and \textit{Swift} observations.}
		\begin{tabular}{llllll}
			System                & Observatory    & Filter & Exposure time (s) & Observation ID           & Observation date         \\   \hline \hline
			NGC 1614N/S   & \textit{Swift}          & UVM2   & 3271.5            & 00046270001; 00046270002 & 2012 May 10; 2012 Jul 6 \\
			AM1054-325/ESO 376-28 & \textit{GALEX} & FUV    & 108               & 6386924688810967040      & 2007 Feb 3               \\
			NGC 2992/3     & \textit{GALEX} & FUV    & 1045.5            & 2485918962089459712      & 2005 Feb 12              \\
			MCG-03-13-063         & \textit{GALEX}          & FUV    & 204               & 6381260060739239936      & 2006 Jan 8               \\			
			NGC 6872              & \textit{GALEX}          & FUV    & 3371.3            & 2505622210593423360      & 2006 Jun 29              \\
			NGC 3256E/W   & \textit{Swift} & UVM2   & 1867.2            & 00049720003; 00049720012 & 2013 Sep 15; 2014 Sep 18 \\
			NGC 1487E/W   & \textit{GALEX} & FUV    & 108               & 6385833989382340608      & 2007 Jan 5               \\ \hline

		\end{tabular}

		\label{UV}
	\end{table*}
	
	\subsection{\textit{Swift}}
	NGC 3256 and NGC 1614 do not have FUV exposures with \textit{GALEX}; instead, we have opted to use NUV data
	from \textit{Swift}, downloaded from the MAST archive.
	The \textit{Swift} satellite operates the 30 cm UltraViolet-Optical Telescope (UVOT), with a FOV of 17 x 17 arcmin. Images were taken in
	2 $\times$ 2 binned mode, with a pixel scale of 1'' per pixel. The UVOT CCD operates as a photon counter, which is susceptible to coincidence loss if two or 
	more photons arrive within a single frame. The effect of this scales with brightness, and past analysis has shown that coincidence loss becomes
	greater than 1\% when the count rate is greater than 0.007 counts sec$^{-1}$ pixel$^{-1}$; for 2 x 2 binned images, as we use, the count rate
	threshold is then 0.028 counts sec$^{-1}$ pixel$^{-1}$. The faint tails of these two systems, NGC 3256 and NGC 1614, fall well below this threshold,
	and we can discount effects from coincidence loss.
	
	\textit{Swift} offers three UV filters, \textit{uvw1}, \textit{uvw2}, and \textit{uvm2}. We use the \textit{uvm2} filter ($\lambda_{\text{eff}} = 2221 \AA$)
	as the other two filters have extended red tails which leak optical light into the UV. Images were co-added together using the HEASOFT software package.
	
	Photometry was performed using the previously mentioned procedure for \textit{GALEX} imaging. Extinction coefficients were
	taken from \cite{roming_08} for the \textit{uvm2} filter.

	\subsection{Tail definition} \label{sec:def}
	Regions defined as tidal debris are taken from \cite{mullan_11}.
	The ``in-tail'' and ``out-of-tail'' regions were defined using images
	taken with the WFPC2 \V\ filter.
	Images were smoothed with a Gaussian
	kernel at 5 - 7 pixels FWHM, and a contiguous region one count above background was defined as ``in-tail'', using SAO DS9; all other regions were
	defined as ``out-of-tail''. In the cases of
	NGC 1487W, AM1054-325, ESO 376-28, and NGC 6872, the centre of the galaxy is imaged as well. The boundary between 
	the centre and the tail is found where the radial light profile changes in scale length.

	\subsection{Cluster Detection} \label{detect}
	Objects were detected using the DAOPHOT.DAOFIND \citep{stetson_87} task in IRAF. Selection criteria for the initial WFPC2 cluster list required 2 counts per object
	in both \V\ and \I, a signal-to-noise (S/N) ratio of at least 3, error in F606W less than 0.25 mag, and detections in at least
	one dither position. For our ACS/WFC3 imaging, we require an S/N of at least 3 in both \U\ and \B/\BA. Objects
	had to be detected in all four filters. We further excluded objects which were fit to our simple stellar population cluster models \citep{marigo_08} with a $\chi^2 > 3$.
	
	Magnitude and colour cuts were applied to our source catalogue to separate potential clusters from contaminant stars and background
	galaxies. Objects which met these requirements are defined as Star Cluster Candidates (SCCs). We apply a magnitude cut of $M_V < -8.5$ ($M_{F606W} < -8.6$)
	designed to eliminate individual stars. Magnitude cuts between $-8 < M_V < -9$ are commonly used in studies of star clusters \citep{konstantopoulos_10}. 
	\cite{whitmore_10} found that even at fainter magnitudes, down to $M_V = -7$, more than 60\% of detected objects were clusters as opposed to individual stars, adding confidence to our selection criteria. 
	In addition to discriminating against non-clusters, a magnitude cut allows a uniform measurement standard across our systems, most of which have nearly complete samples
	of SCCs down to the M$_V$ = -8.5 cutoff, as shown in Section \ref{complete}. 
	A colour cut of $V - I < 2.0$ (\V\ - \I\ < 1.4) is added as well; this will still allow for old globular clusters which have reddened as they 
	evolve, while eliminating individual stars.


	\subsection{Completeness} \label{complete}
\begin{figure*}
		\centering
		\includegraphics[width=0.8\linewidth]{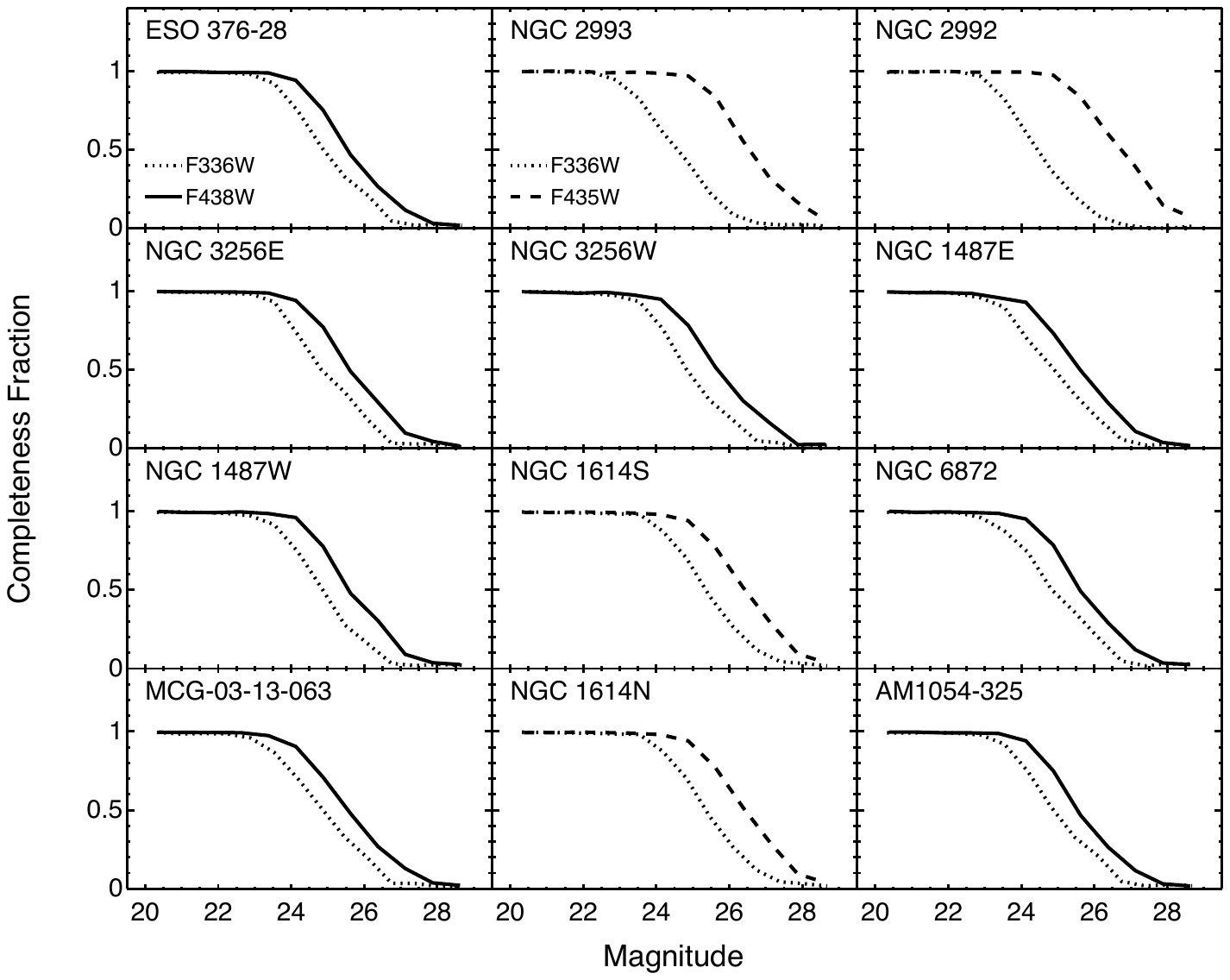}
		\vspace{-0.8cm}
		\caption{Completeness data for our WFC3 F336W/F435W and ACS F438W imaging. Solid curves
			represent data for F438W, dotted curves represent data for F336W, and dashed curves
			represent data for F435W.}
		\label{fig:complete}
	\end{figure*}

	Completeness curves from \cite{mullan_11} show that our WFPC2 data are on average
	50\% complete at $m_\text{\V} \approx 25.5$ and $m_\text{\I} \approx 24.5$.
	To measure the completeness of our WFC3 and ACS imaging, we perform a similiar 
	analysis as in \cite{mullan_11}; we add 10,000 fake stars to each individual image,
	100 at a time, with DAOPHOT.ADDSTAR \citep{stetson_87}, and calculate how many are recovered
	above a 3$\sigma$ limit. Our completeness curves are shown in Figure \ref{fig:complete}.
	For systems AM1054-325, ESO 376-28, NGC 3256, NGC 1487, MCG-03-13-063, and NGC 6872, we
	are complete at 50\% at $m_\text{\B} = 25.5$ and $m_\text{\U} = 24.8$. NGC 2992 and NGC 2993
	are 50\% complete at $m_\text{\BA} = 26.5$ and $m_\text{\U} = 24.5$. NGC 1614, observed
	with program GO-14066 and GO-10592 is complete at $m_\text{\BA} = 26.5$ and $m_\text{\U} = 25.4$.
	
	\section{Results} \label{sec:3_2}
	\subsection{Colour-colour diagrams}
\begin{figure*}
	\centering
	\includegraphics[width=0.9\linewidth]{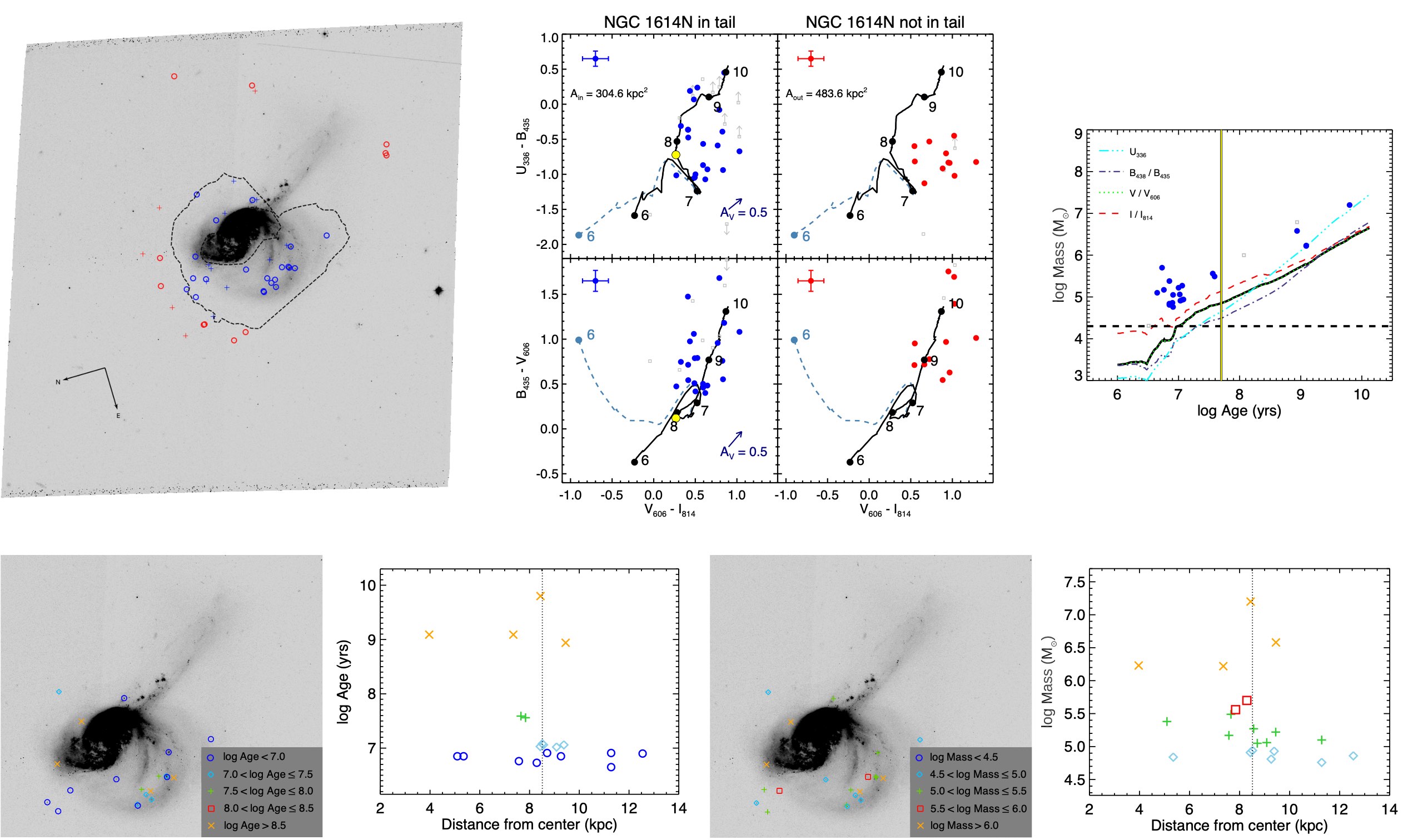}
	\caption{\textit{Top left}: \textit{HST} \BA-band image of NGC 1614N, with the tail region outlined with a black, dashed curve. 
	In-tail SCCs are shown as blue circles, while out-of-tail SCCs are red circles.
	Non-SCC detected objects are shown as crosses. \textit{Top middle}: Colour-colour diagrams for in-tail
	and out-of-tail sources plotted against stellar evolutionary tracks from \protect\cite{marigo_08} in solid black; the dashed light blue
	line is the same track with nebular continuum and emission lines added. The yellow circle indicates the age of the merger.
	Blue/red circles represent SCCs for in-tail/out-of-tail sources, and gray boxes are non-SCC detected objects. The total area enclosed in each
	in-tail or out-of-tail region in kpc$^2$ is indicated on the upper left, along with median error bars. \textit{Top right}: Ages and masses for detected 
	objects. SCCs are shown in blue, non-SCCs in gray.
	The solid black curve is our magnitude limit of M$_V$ = -8.5, while the red, green, purple,
	and cyan curves show 50\% completeness limits. The vertical yellow line 
	marks the interaction age of the system, and the horizontal dashed line marks our mass
	cut-off of log Mass = 4.3 for our CFE determinations. \textit{Bottom row}: age and mass distributions for our SCCs. A vertical dashed line marks the median
	distance from the centre. We find 21 in-tail SCCs, and 11 out-of-tail SCCs for NGC 1614N. The tail curls to the North from the East, with young clusters scattered throughout the region.}
	\label{fig:1614N_all}
\end{figure*}

\begin{figure*}
	\centering
	\includegraphics[width=0.9\linewidth]{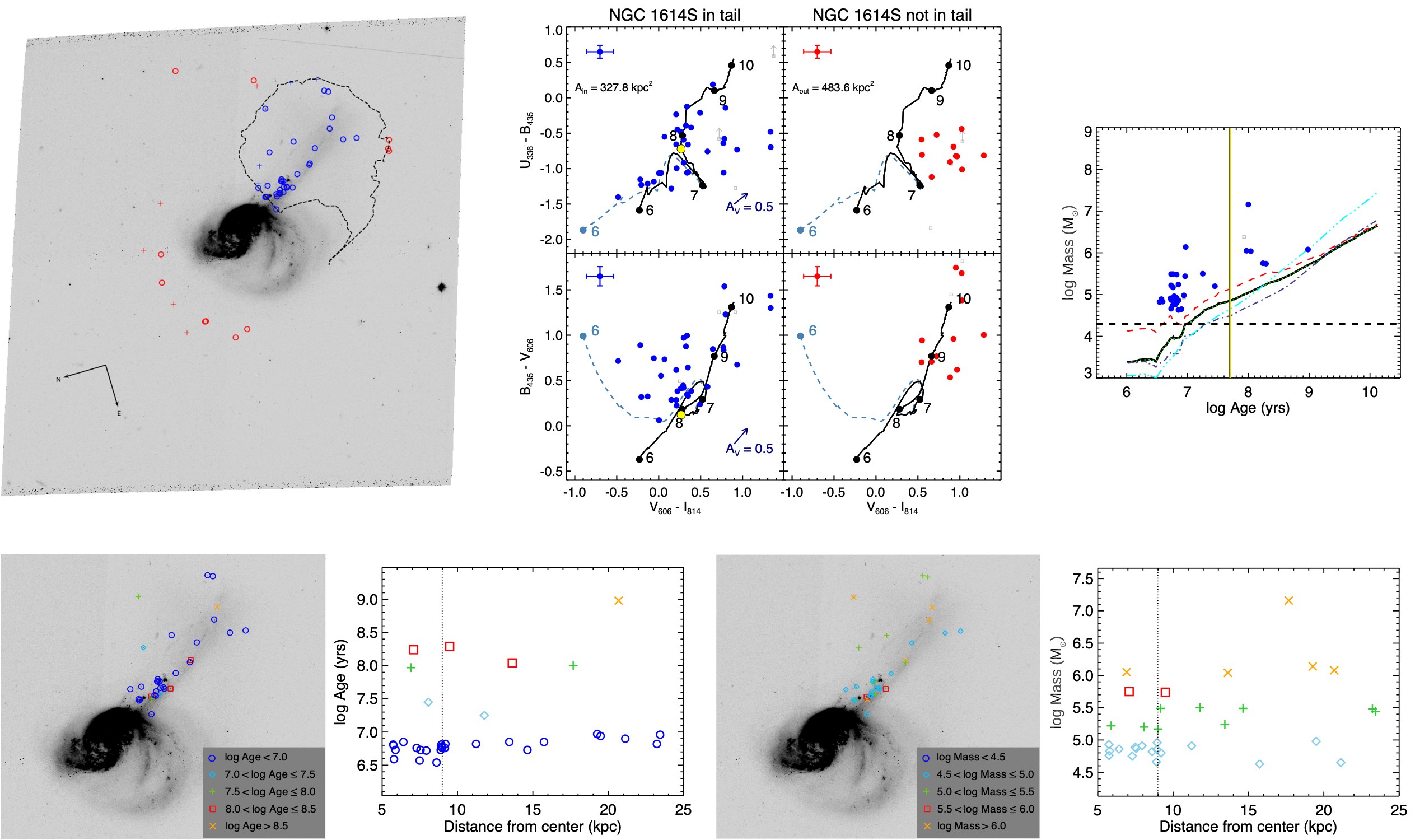}
	\caption{Same as \ref{fig:1614N_all}, but for NGC 1614S. We find 33 in-tail SCCs, and 11 out-of-tail SCCs. This system has a 
		statistically significant excess cluster density above $3\sigma$. The tail extends to the south from the center of the 
		merging system. Young objects are found throughout the tail, similar to NGC 1614N.}
	\label{fig:1614S_all}
\end{figure*}

\begin{figure*}
	\centering
	\includegraphics[width=0.9\linewidth]{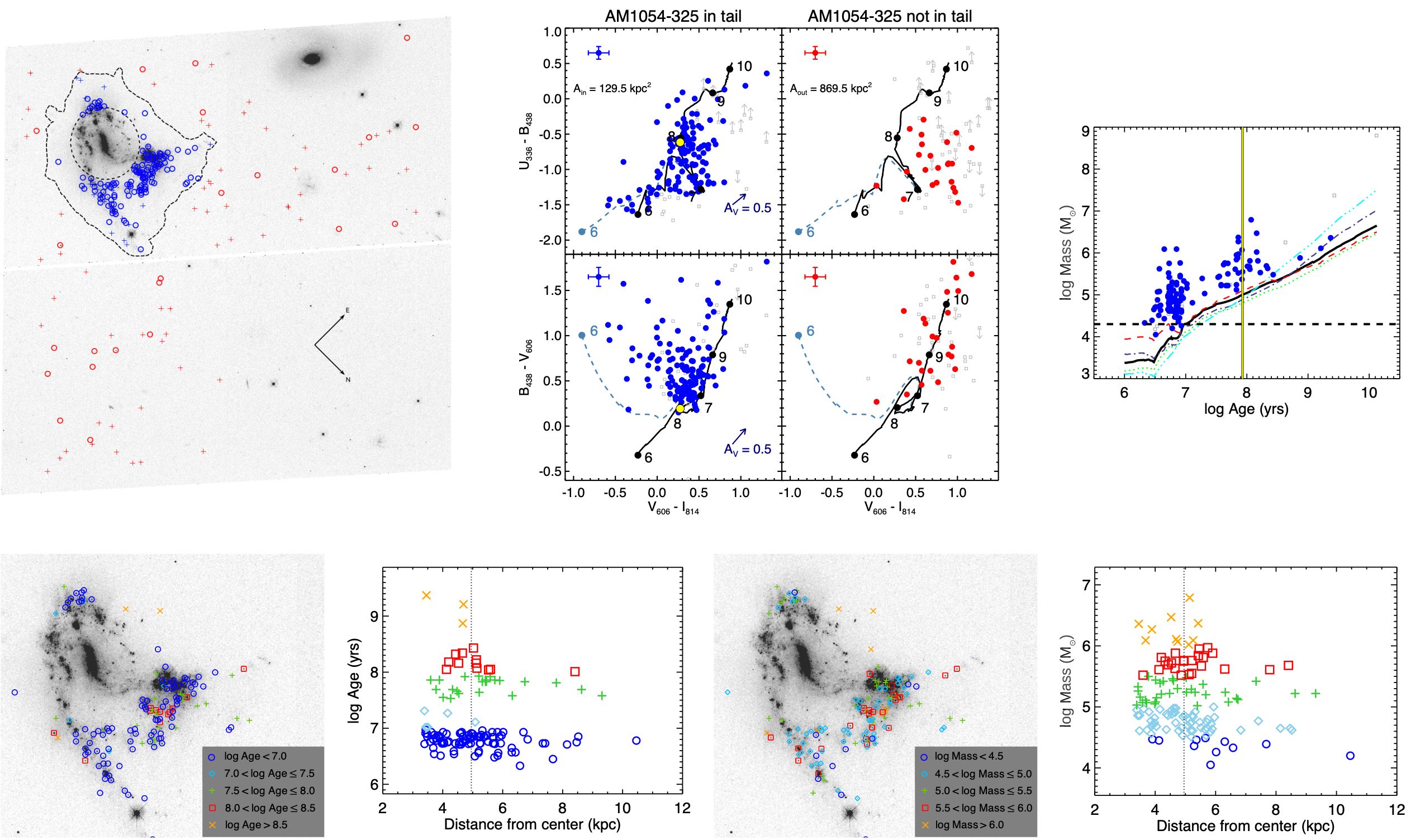}
	\caption{Same as \ref{fig:1614N_all}, but for AM1054-325. We find 135 in-tail SCCs, 
		and 23 out-of-tail SCCs. This system has a statistically significant excess cluster 
		density above $3\sigma$. A tidal dwarf is directly North of the center of the galaxy, 
		while the tidal tail extends Northward from the Western edge. Many SCCs show signs 
		of emission lines, indicated by their large B – V values. Its interacting partner can be 
		seen at the top right of the image.}
	\label{fig:AM1054_all}
\end{figure*}

\begin{figure*}
	\centering
	\includegraphics[width=0.9\linewidth]{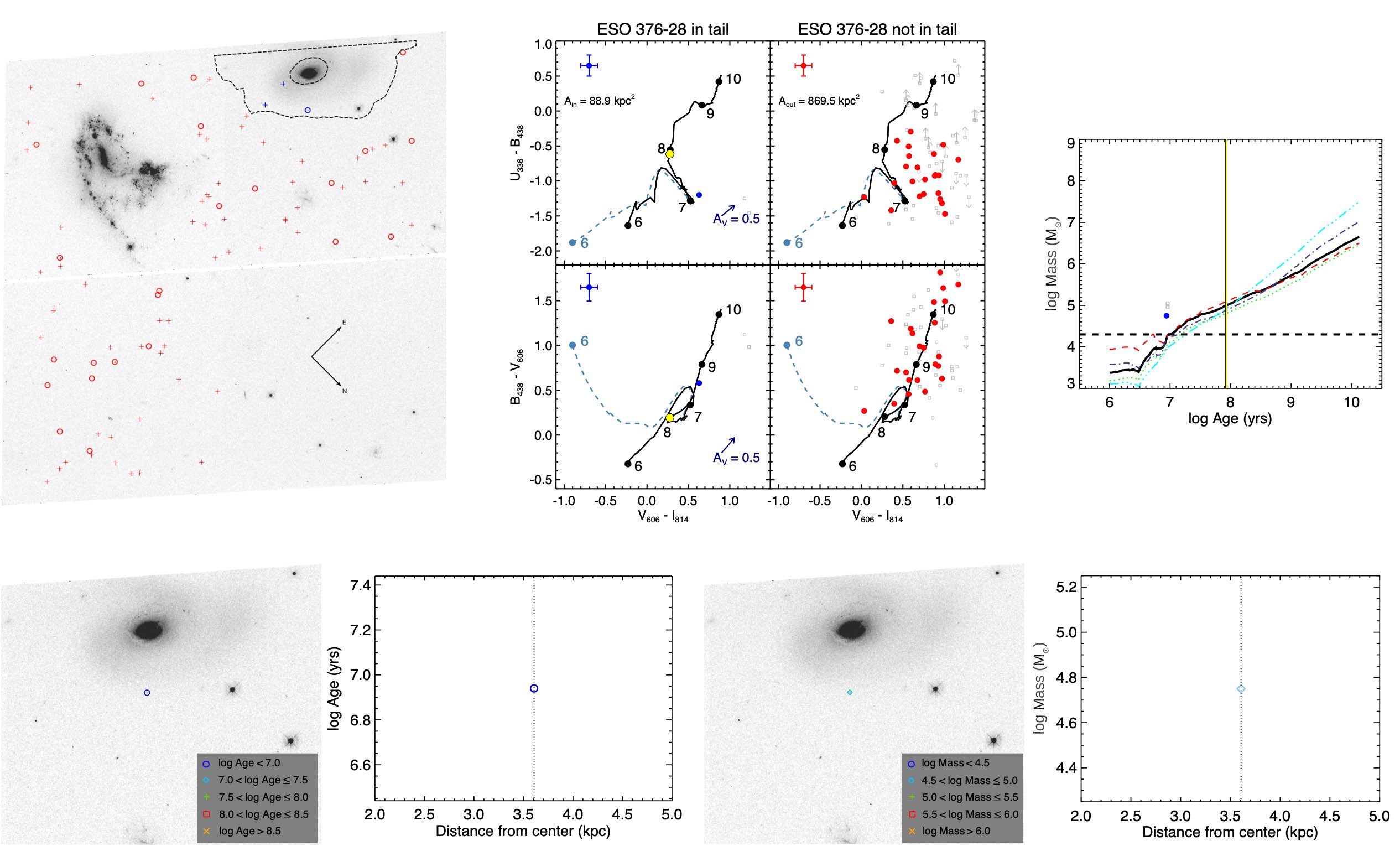}
	\caption{Same as \ref{fig:1614N_all}, but for ESO 376-28. We find 1 in-tail SCC, and 23 out-of-tail SCCs. Little structure is seen in this galaxy.}
	\label{fig:ESO_all}
\end{figure*}

\begin{figure*}
	\centering
	\includegraphics[width=0.9\linewidth]{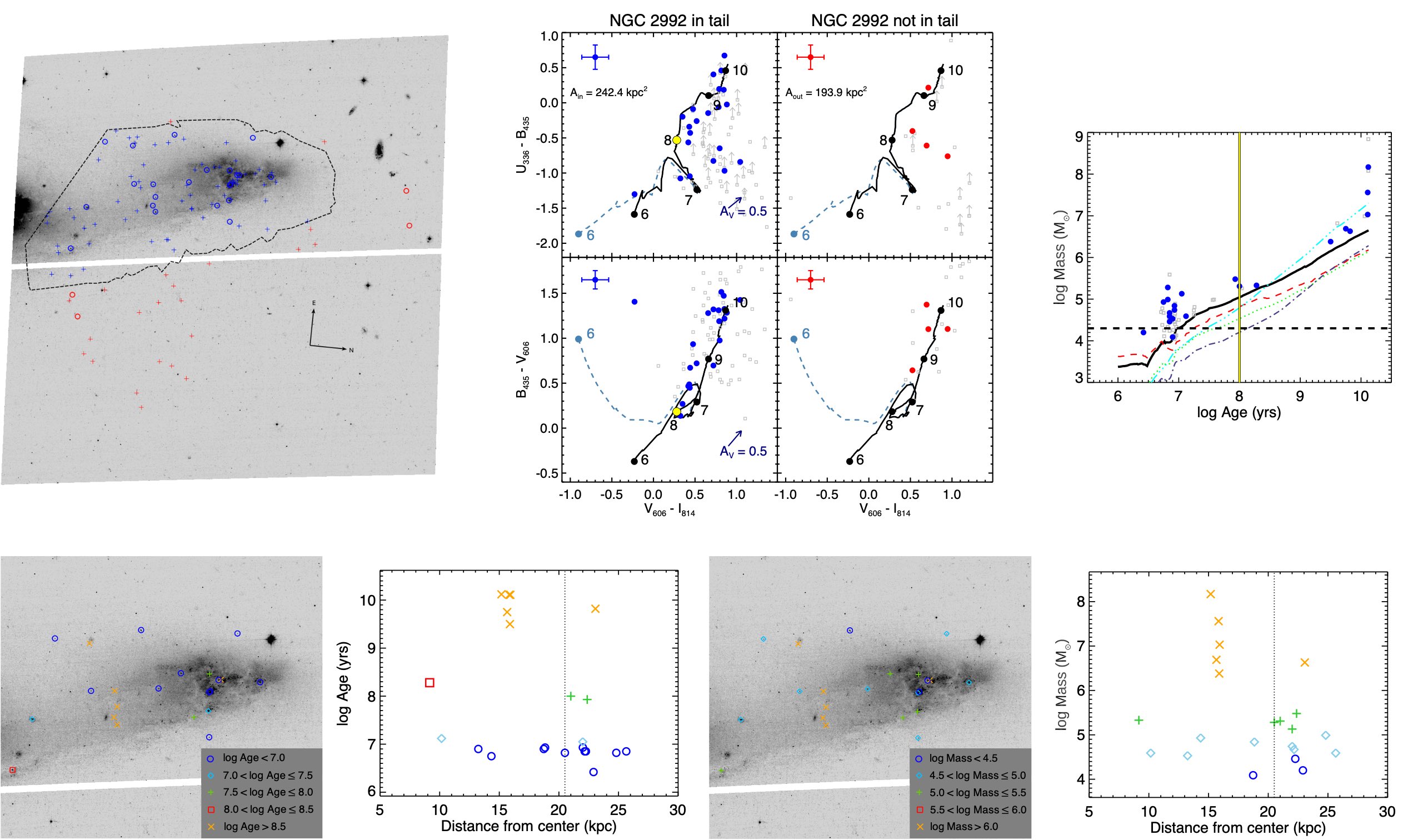}
	\caption{Same as \ref{fig:1614N_all}, but for NGC 2992. We find 22 in-tail SCCs, and 4
		 out-of-tail SCCs. This system has a statistically significant excess cluster density 
		 above $3\sigma$. We have targeted the tidal dwarf, with the Northern edge of the 
		 galaxy NGC 2992 shown on the left. We do not include sources from the galaxy itself.}
	\label{fig:2992_all}
\end{figure*}

\begin{figure*}
	\centering
	\includegraphics[width=0.9\linewidth]{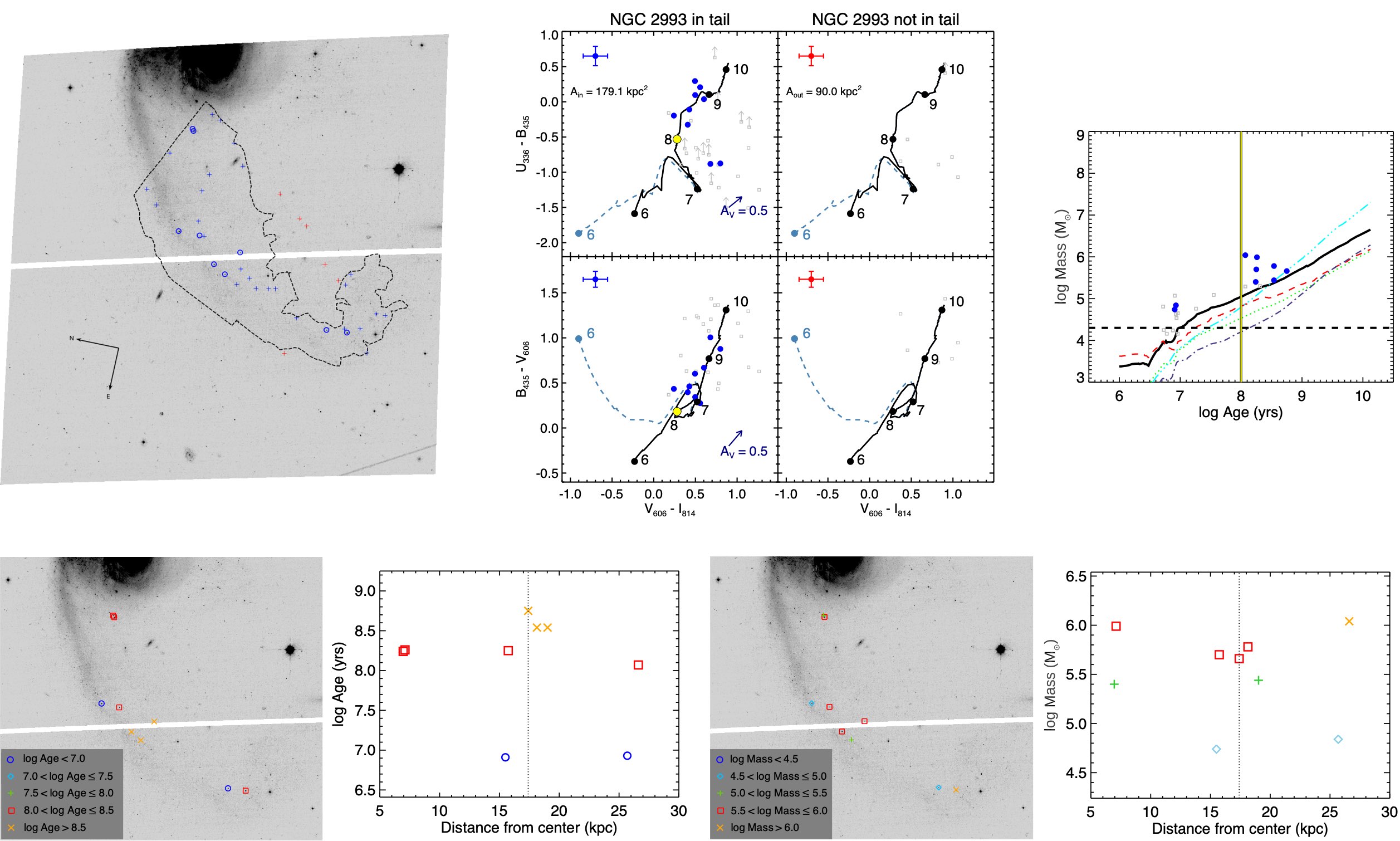}
	\caption{Same as \ref{fig:1614N_all}, but for NGC 2993. We find 9 in-tail SCCs, and 0
		 out-of-tail SCCs. We capture the tidal tail of NGC 2993 (seen at the top of the 
		 image). Most of the SCCs in the tail have ages comparable to the interaction 
		 age of the NGC 2992/3 system.}
	\label{fig:2993_all}
\end{figure*}

\begin{figure*}
	\centering
	\includegraphics[width=0.9\linewidth]{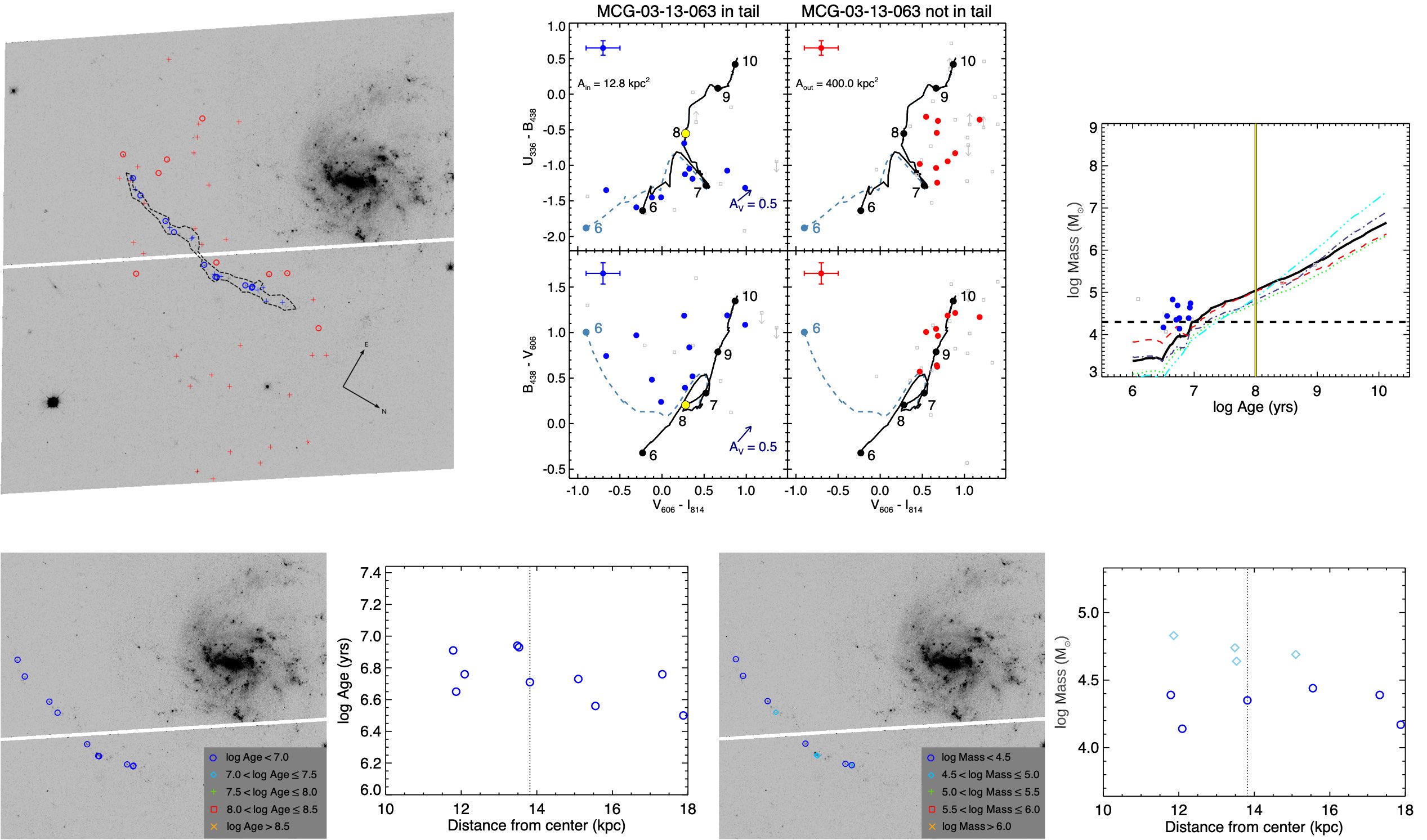}
	\caption{Same as \ref{fig:1614N_all}, but for MCG-03-13-063. We find 10 in-tail SCCs, 
		and 9 out-of-tail SCCs. This system has a statistically significant excess cluster 
		density above $3\sigma$.  An extended, thin tail is seen, produced from an unseen
		companion. All the SCCs in the tail are < 10 Myr.}
	\label{fig:MCG_all}
\end{figure*}

\begin{figure*}
	\centering
	\includegraphics[width=0.9\linewidth]{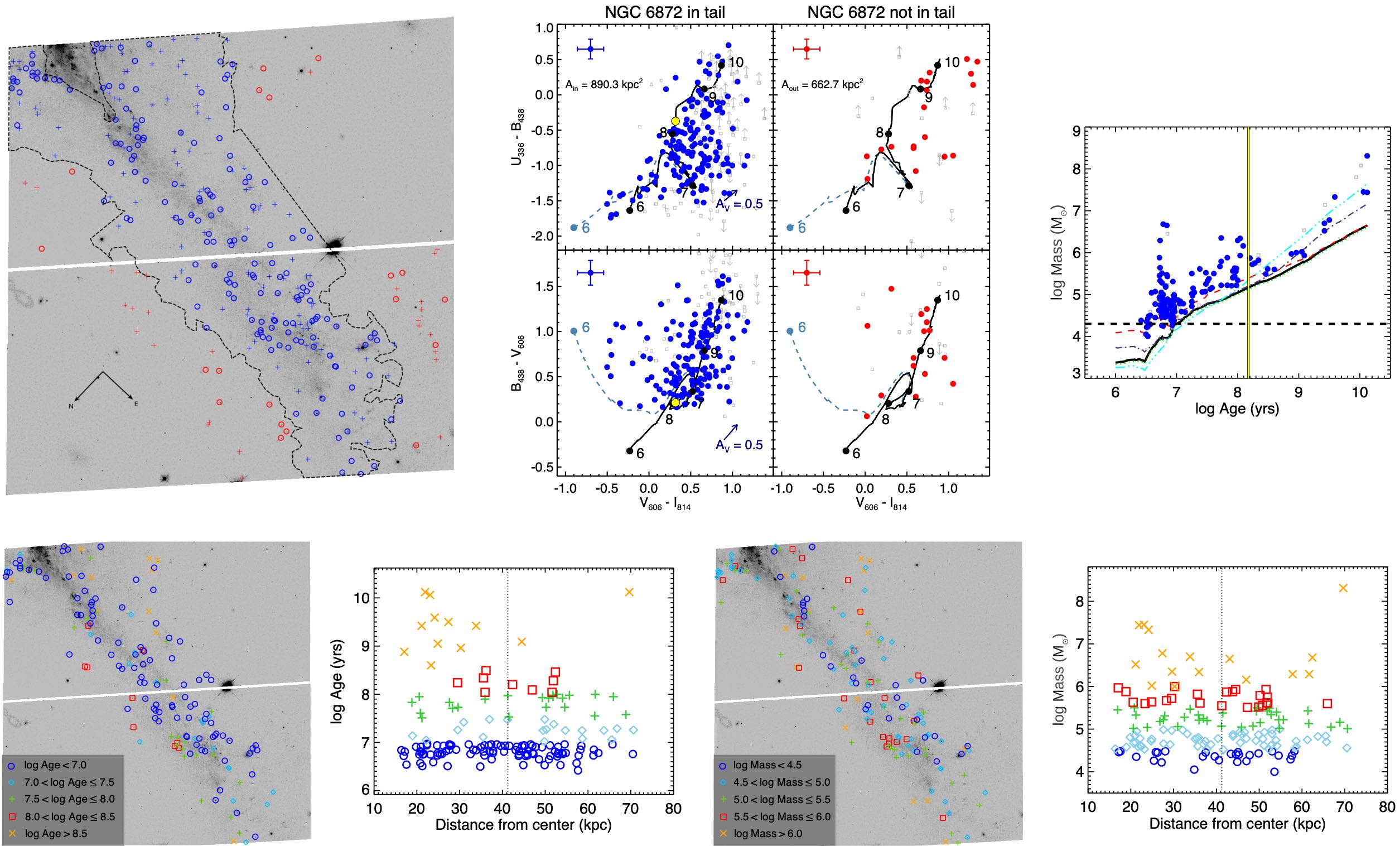}
	\caption{Same as \ref{fig:1614N_all}, but for NGC 6872. We find 158 in-tail SCCs, and 
		19 out-of-tail SCCs. This system has a statistically significant excess cluster density 
		above $3\sigma$. This is Eastern tidal tail of NGC 6872, which stretches out to 
		70 kpc; the center of the galaxy lies to the West. A range of ages are seen, with 
		young SCCs spread out along the length of the tail.}
	\label{fig:6872_all}
\end{figure*}

\begin{figure*}
	\centering
	\includegraphics[width=0.9\linewidth]{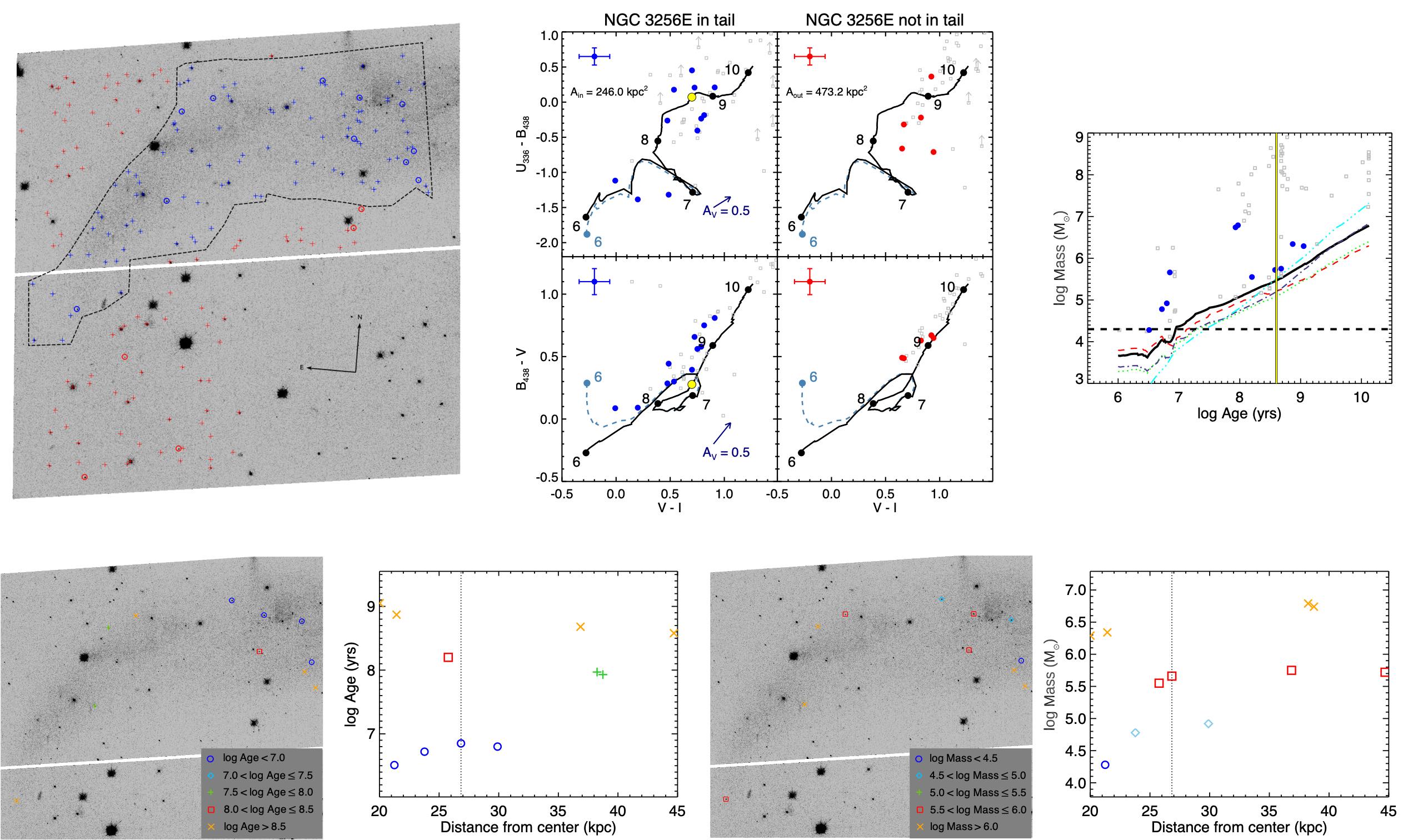}
	\caption{Same as \ref{fig:1614N_all}, but for NGC 3256E. We find 11 in-tail SCCs, 
		and 5 out-of-tail SCCs. Emission line clusters do not clearly stand out, as the 
		NGC 3256 system was imaged with the F555W filter, which weakly covers the 
		H$\alpha$ line. Despite the old age of the interaction, we see a few SCCs with 
		young ages.}
	\label{fig:3256E_all}
\end{figure*}

\begin{figure*}
	\centering
	\includegraphics[width=0.9\linewidth]{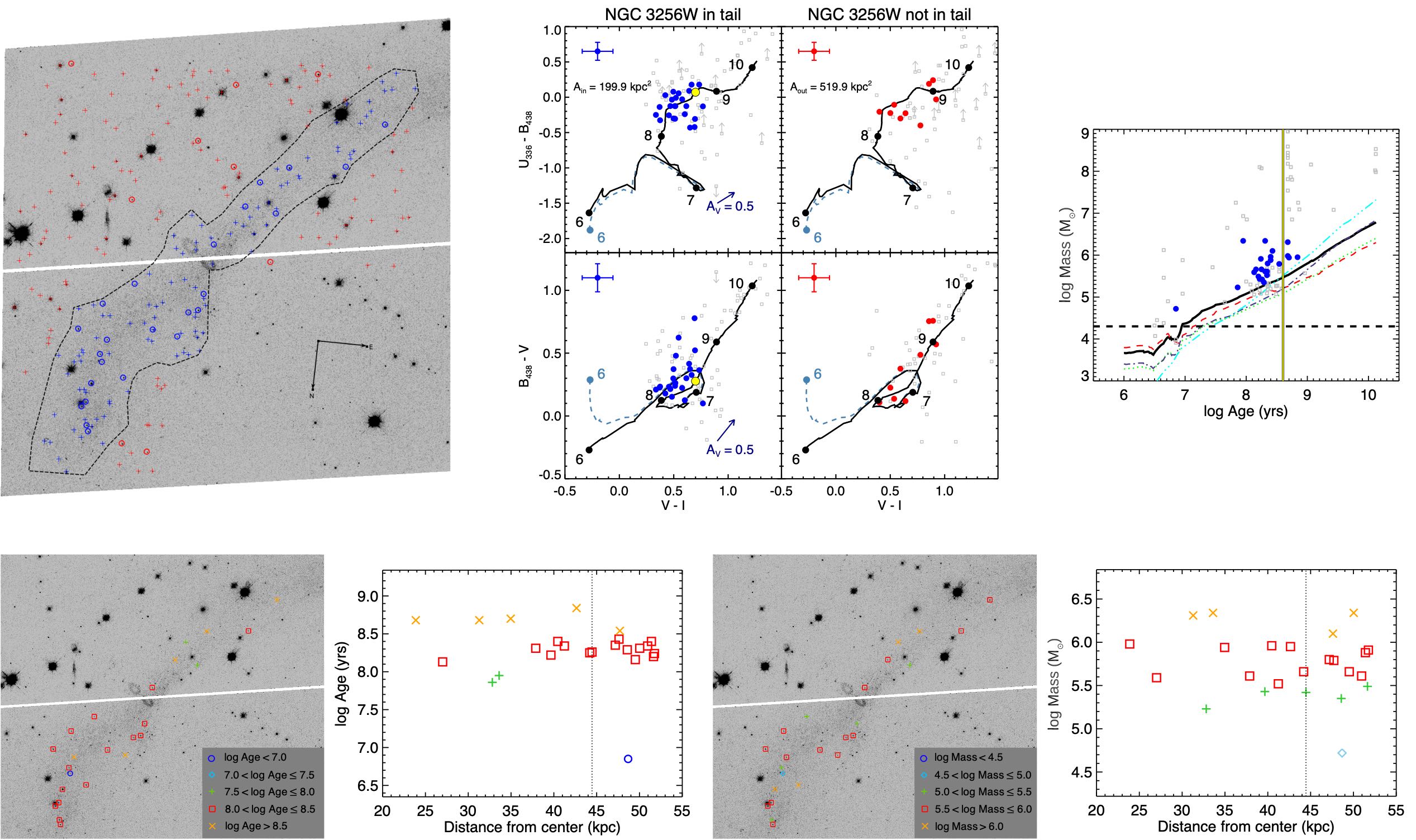}
	\caption{Same as \ref{fig:1614N_all}, but for NGC 3256W. We find 24 in-tail 
		SCCs, and 9 out-of-tail SCCs. This system has a statistically significant excess 
		cluster density above $3\sigma$. Many of the SCCs are clustered around the 
		interaction age of the merger, suggesting they formed from the interaction.}
	\label{fig:3256W_all}
\end{figure*}

\begin{figure*}
	\centering
	\includegraphics[width=0.9\linewidth]{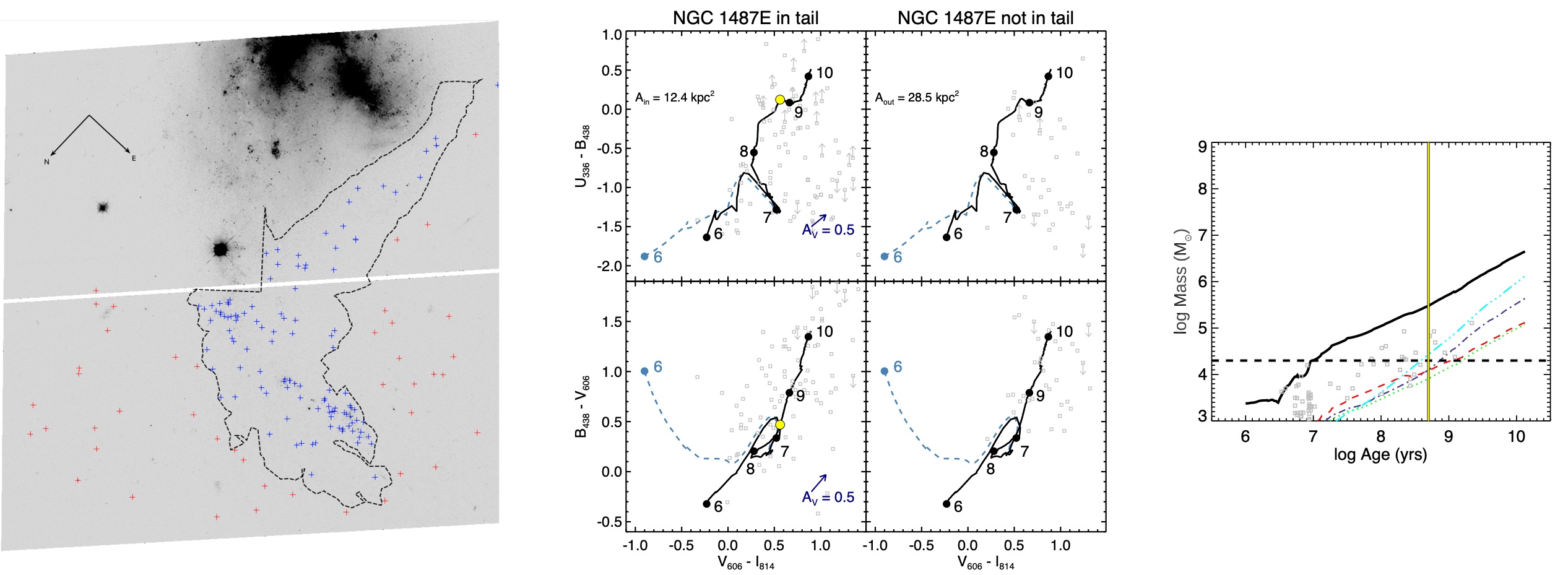}
	\caption{Same as \ref{fig:1614N_all}, but for NGC 1487E. We find 0 in-tail SCCs, and 0 out-of-tail SCCs. Note that as NGC 1487E has no detected
		SCCs, age and mass distribution figures are not included. Although no SCCs are 
		found in this system, a large number of non-SCC objects are detected, which do 
		not make the $M_V < -8.6$ cutoff. This suggests the presence of low-luminosity,
		low-mass objects.}
	\label{fig:1487E_all}
\end{figure*}

\begin{figure*}
	\centering
	\includegraphics[width=0.9\linewidth]{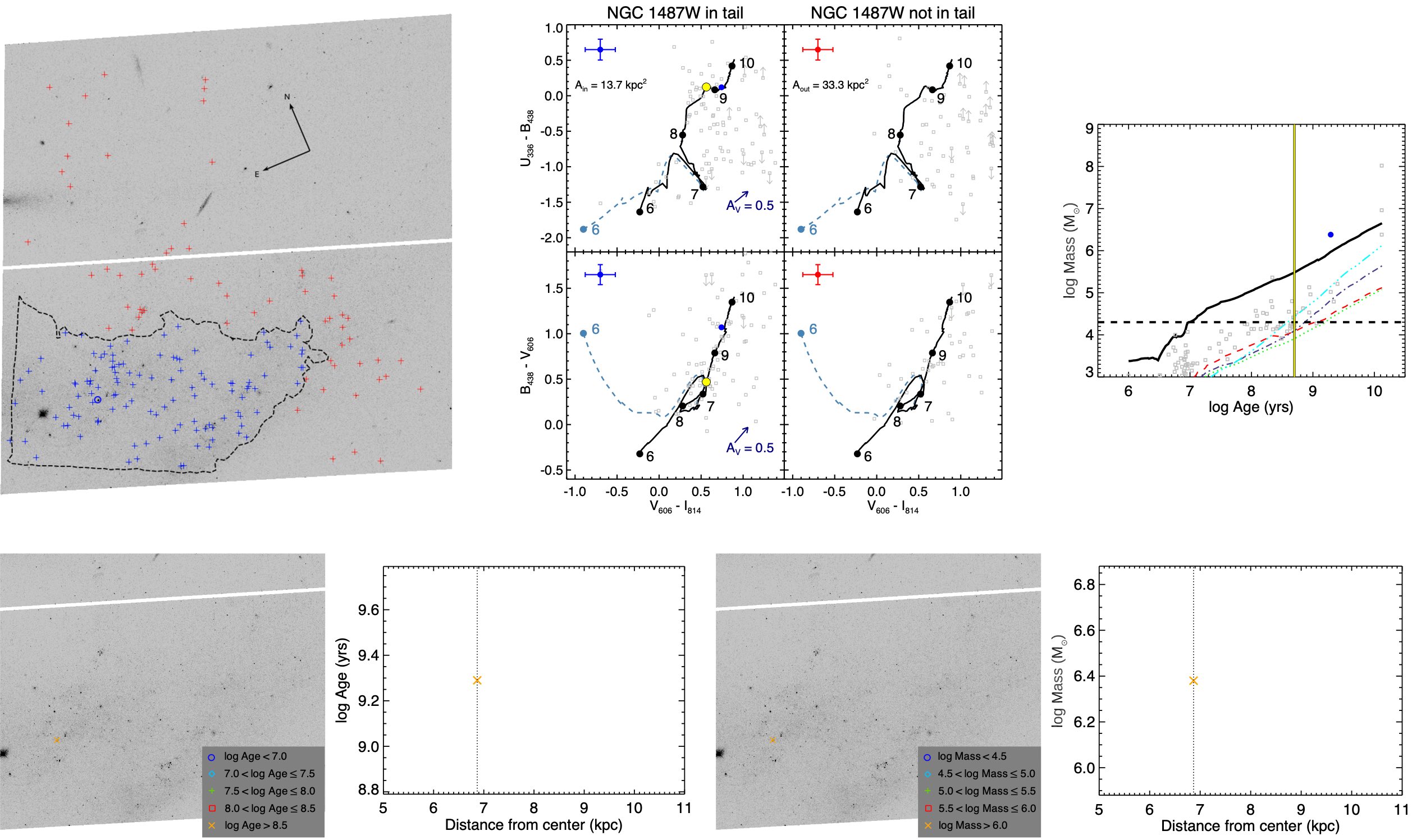}
	\caption{Same as \ref{fig:1614N_all}, but for NGC 1487W. We find 1 in-tail SCC, 
		and 0 out-of-tail SCCs. Like NGC 1487E, we again find a number of non-SCC 
		objects in the tail, suggesting low-mass objects present in the tail.}
	\label{fig:1487W_all}
\end{figure*}

	In Figures \ref{fig:1614N_all}-\ref{fig:1487W_all} we show the \U\ - \B\ vs \V\ - \I\ colour-colour diagrams for all our observed systems. Systems are ordered in reference to the age of the tidal debris (see Table \ref{table:data}).
	The interaction age of the merger is plotted as a yellow circle. Objects which fulfill our SCC criteria in Section \ref{detect} are shown
	as dark blue circles. For completeness, we show objects which do not meet our criteria as gray boxes, with arrows indicating upper and lower
	limits. 
	
	Data are plotted against simple stellar population (SSP) models
	from \cite{marigo_08}, with logarithmic ages overplotted on the evolutionary track. We use a Salpeter IMF; however, the choice
	of IMF is negligible in determining ages, and the tracks are also consistent
	expectations for a Chabrier \citep{chabrier_01} or Kroupa \citep{kroupa_01} IMF. We corrected for foreground Galactic extinction using an $R_\lambda = A_\lambda/E(B-V)$ 
	reddening law, with $R_V = 3.1$, with data from \cite{schlafly_11}. Internal cluster extinction is not corrected for when plotting our colour-colour diagrams. However, this is taken into account when
	SED fitting our clusters (Section \ref{age_mass}). We also include a reddening arrow in the lower 
	right hand corner of our plots for A$_V$ = 0.5.

	The area inside and outside the tail was calculated using SAO DS9 regions of the tail, as defined in Section \ref{sec:def}. We subtract the cluster density 
	outside the tail ($N^{\textrm{SCC}}_{\textrm{out}} / A_\textrm{out}$) from inside the tail
	($N^{\textrm{SCC}}_{\textrm{in}} / A_\textrm{in}$) to determine the excess cluster density, $\Sigma_\textrm{SCC}$.
	Errors are determined from Poisson statistics. Half of our tails show excesses above the
	$3\sigma$ level, indicating significant amounts of SCCs. It is important to note that while some systems
	may not contain significant numbers of SCCs, some individual objects in tails may still be real star clusters. 
	The data are shown in Table \ref{table:info}.

	\begin{table*}
			\caption{Cluster excesses.}
			\begin{center}
	\begin{tabular}{llllllllll}
		System &  $N^{\text{SCC}}_{\text{in}}$ & $N^{\text{SCC}}_{\text{out}}$ & A$_{\text{in}}$ & 	A$_{\text{out}}$ & Density$_{\text{in}}$ & Density$_{\text{out}}$ & $\Sigma_{SCC}$   \\ 
		                                         &         &          & (kpc$^2$) & (kpc$^2$) & (kpc$^{-2}$) & (kpc$^{-2}$) & (kpc$^{-2}$)       \\ \hline \hline
		NGC 1614N                             & 21               & 11                & 304.6          & 483.6           & 0.069                & 0.023                 & $0.046 \pm 0.017$  \\	
		NGC 1614S                             & 33               & 11                & 327.8          & 483.6           & 0.101                & 0.023                 & $0.078 \pm 0.019^*$  \\
		AM1054-325                              & 135              & 23                & 129.5          & 869.5           & 1.042                & 0.026                 & $1.016 \pm 0.090^*$ \\
		ESO 376-28        & 1                & 23                & 88.9           & 869.5           & 0.011                & 0.026                 & $-0.015 \pm 0.013$ \\
		NGC 2992                              & 22               & 4                 & 242.4          & 193.9           & 0.091                & 0.021                 & $0.070 \pm 0.022^*$  \\		
		NGC 2993                                 & 9                & 0                 & 179.1          & 90.0            & 0.050                & 0                     & $0.050 \pm 0.017$  \\
		MCG-03-13-063                           & 10               & 9                 & 12.8          & 400.0           & 0.781                & 0.023                 & $0.76\pm 0.25^*$  \\
		NGC 6872                             & 158              & 19                & 890.3          & 662.7           & 0.179                & 0.029                 & $0.150 \pm 0.016^*$  \\
		NGC 3256E                              & 11               & 5                & 246.0          & 473.2           & 0.045                & 0.011                 & $0.034 \pm 0.014$  \\
		NGC 3256W                              & 24               & 9                & 199.9          & 519.9           & 0.120                & 0.017                 & $0.103 \pm 0.025^*$  \\
		NGC 1487E                                               & 0                & 0                 & 12.4           & 28.5            & 0                    & 0                     & 0                  \\
		NGC 1487W                                                & 1                & 0                 & 13.7           & 33.3            & 0.073                & 0                     & $0.073 \pm 0.073$  \\ \hline

	\end{tabular}
\end{center}
\raggedright
$^*$Signifies excess at 3$\sigma$ or above.
	\label{table:info}
\end{table*}
	
	We perform Kolmogorov--Smirnov (KS) tests on the distributions of \U\ - \B, \B\ - \V, and \V\ - \I\ 
	colours between in-tail and out-of-tail objects for all detected objects and for SCCs only. This is another way to determine the likelihood 
	that objects within the tail are unique and independent from those outside the tail region, in addition to measuring cluster excess. Recorded 
	\textit{p}-values are shown in Table \ref{table:KS}. 
	Our \V\ - \I\ colour is most useful in discriminating between tail and non-tail objects. Of the eight systems which contain
	enough SCCs for a KS test, five of them show \textit{p}-values less than 0.04, indicating the data are drawn from independent distributions.
	
	\begin{table*}
		\caption{KS test results for our photometric data. We include results for all of our sources on the left, and 
	SCC objects only on the right.}
		\begin{tabular}{lllllllllll}
			\multicolumn{1}{c}{} &
			\multicolumn{5}{c}{All} &
			\multicolumn{5}{c}{SCC only}\\
			\cmidrule(lr){2-6} \cmidrule(lr){7-11}
			
			System        & N$_{in}$ & N$_{out}$ & KS$_{U - B}$   & KS$_{B - V}$    & KS$_{V - I}$   & N$_{in}$ & N$_{out}$ & KS$_{U - B}$   & KS$_{B - V}$    & KS$_{V - I}$ \\ \hline \hline
			NGC	1614N	&	26	&	14	&	0.025	&	0.020	&	6.6 $\times 10^{-3}$	&   21	&	11	&	0.17	&	0.19	&	7.2 $\times 10^{-3}$		\\
			NGC	1614S	&	40	&	14	&	0.068	&	2.0 $\times 10^{-3}$	&	5.852 $\times 10^{-5}$	&   33	&	11	&	0.65	&	0.045	&	1.2$ \times 10^{-4}$  \\		
			AM1054-325	&	172	&	51	&	0.40	&	6.9 $\times 10^{-6}$	&	6.6 $\times 10^{-18}$	&   135	&	23	&	0.166	&	4.5 $\times 10^{-3}$  & $2.0 \times 10^{-8}$ \\
			ESO	376-28	&	3	&	51	&	N/A	&	N/A	&	N/A	& 1	&	23	&	N/A	&	N/A	&	N/A		\\					
			NGC	2992	&	61	&	15	&	0.058	&	0.42	&	0.75	&   22	&	4	&	0.73	&	0.87	&	0.50		\\			
			NGC	2993	&	23	&	5	&	0.19	&	0.018	&	0.34   &   9	&	0	&   N/A	&	N/A	&	N/A		\\
			MCG-03-13-063	&	19	&	24	&	2.3 $\times 10^{-3}$	&	0.73	&	1.8 $\times 10^{-3}$	&   10	&	9	&	2.1 $\times 10^{-3}$	&	0.35	&	1.7 $\times 10^{-3}$		\\
			NGC	6872	&	205	&	29	&	1.6 $\times 10^{-3}$	&	0.031	&	0.012  &   158	&	19	&	0.026	&	0.049	&	0.025		\\
			NGC	3256E	&	71	&	53	&	0.74	&	0.021	&	0.061	&   11	&	5	&	0.92	&	0.17	&	0.23		\\	
			NGC	3256W	&	92	&	74	&	0.055	&	8.1 $\times 10^{-3}$	&	6.7 $\times 10^{-4}$&   24	&	9	&	0.97	&	0.33	&	0.18		\\
			NGC	1487E	&	83	&	33	&	0.75	&	0.083	&	4.5 $\times 10^{-3}$	&   0	&	0	&	N/A	&	N/A	&	N/A			\\
			NGC	1487W	&	92	&	44	&	0.39	&	0.36	&	0.46	&   1	&	0	&	N/A	&	N/A	&	N/A			\\ \hline

		\end{tabular}

		\label{table:KS}
	\end{table*}

	\subsection{Addition of nebular flux} \label{sec:nebs}
	
\begin{figure}
		\centering
		\includegraphics[width=1\linewidth]{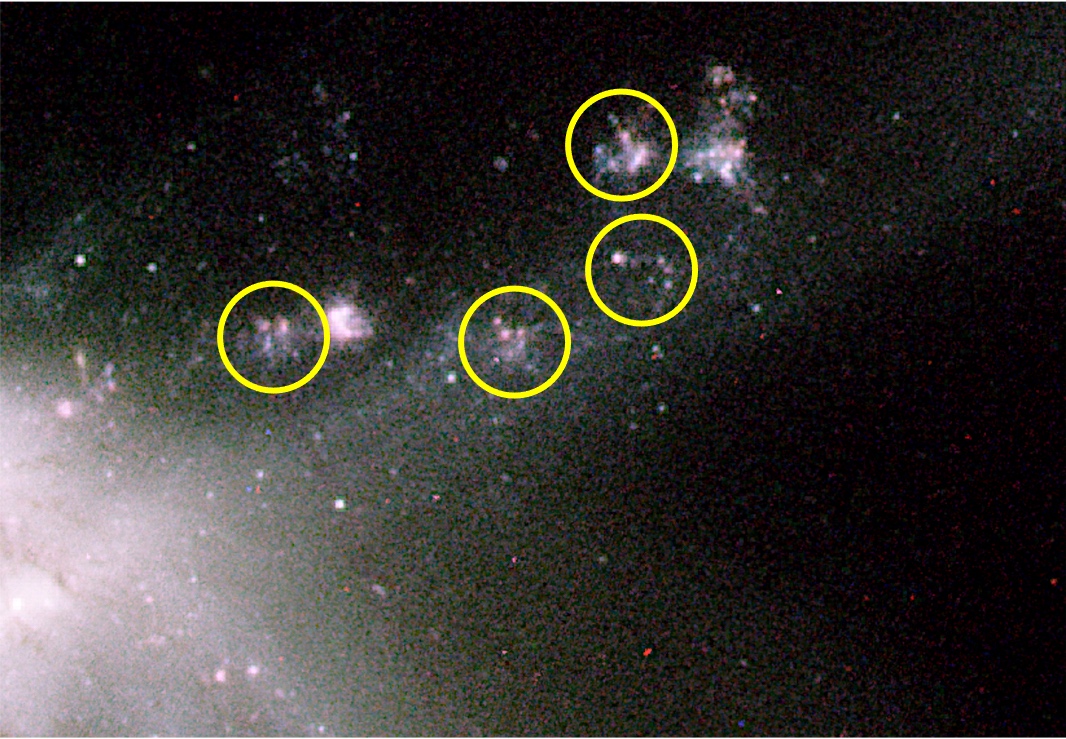}
		\vspace{-0.5cm}
		\caption{Colour image of NGC 1614S. The data through filters F435W (ACS), 
			F606W (WFPC2), F814 (ACS), and F665N (ACS) were stretched with a 
			logarithm and bias/contrast adjustment in CARTA$^a$. Subsequently in GIMP$^b$
			they were assigned the colours blue, yellow-green, red and pink respectively 
			and using a layering schema blended with the screen algorithm \citep{english_17}. 
			Several regions containing SCCs with emission lines are highlighted in yellow, showing
		that star cluster formation is ongoing in this tidal tail.}
		\raggedright
		$^a$CARTA: Cube Analysis and Rendering Tool for Astronomy. \url{https://cartavis.org}
		\newline
		$^b$Gnu Image Manipulation Program. \url{https://www.gimp.org}
		\label{fig:neb}
	\end{figure}

	Clusters less than 10 Myr old can show strong emission lines, as the surrounding hydrogen
	gas from cluster formation is ionized and undergoes recombination. An example of these systems is shown
	in Figure \ref{fig:neb} for NGC 1614. As clusters will expel their hydrogen gas via stellar feedback on timescales of
	several million years \citep{pang_20}, the presence of recombination lines indicates recent star formation.
	
	Our SSP model (black solid line in Figures \ref{fig:1487W_all} - \ref{fig:1614N_all}) does not contain nebular continuum emission, nor flux
	from emission lines; we use Starburst99 \citep{sb99} to calculate contributions from the nebular continuum, as well as the 
	H$\alpha$ and H$\beta$ emission lines. We include the H$\gamma$ line, with the flux ratio of H$\gamma$ / H$\beta$ = 0.47 \citep{osterbrock_89},
	and the O[III] and O[II] lines as well. The oxygen lines are determined from the KPNO International Spectroscopic Survey (KISS) sample
	of nearby, low-mass star forming galaxies \citep{salzer_05}; we take the median log ratio of these lines to the H$\beta$ line, O[III]/H$\beta$ = 0.08 and O[II]/H$\beta$ = 0.56.
	
	The effect of the emission lines depends on the filter. The H$\alpha$, H$\beta$, and O[III] lines are covered by the F606W filter, while the H$\gamma$
	line falls in the middle of the \B\ and \BA\ filters. The O[II] line is at the edge of the F435W and F336W filters. The result is that
	the \V\ filter is most strongly affected by the presence of emission lines, due to the strong H$\alpha$ and H$\beta$ lines, while the effect in the other filters
	is relatively inconsequential. This shifts the \V\ - \I\ colour towards bluer colours in our colour-colour plots. The resulting evolutionary 
	track is degenerate with extinction for ages  $< 10$ Myr for our \U\ - \B\ vs \V\ - \I\ diagrams. The effect is more apparent when we 
	plot \B\ - \V\ vs \V\ - \I; the evolutionary track dramatically swings upward in our plots as the \B\ - \V\ colour trends towards redder values. This effect
	is seen for similar young clusters in interacting systems as well, using the \V\ filter \citep{fedotov_15,fedotov_11,gallagher_10}.
	
	Data for NGC 3256 were taken from \cite{knierman_03}. For NGC 3256W/E, we determined the magnitudes in the F555W and F814 filters, and then
	transformed these to the Johnsons-Cousins system, as in \cite{knierman_03}, using transformations from
	\cite{holtzman_95}. The shape and wavelength boundaries of the F555W filter transmission are noticeably different
	from the \V\ filter in that the H$\alpha$ line falls at the edge of the filter, thus it has a much less of an effect on our model magnitudes. Therefore, on the
	colour-colour diagrams for NGC 3256W/E, there is less of a difference between
	the nebular emission evolutionary tracks and the SSP models than for our other
	systems.
	
	\subsection{Ages and Masses} \label{age_mass}
\begin{figure}
		\centering
		\includegraphics[width=0.9\linewidth]{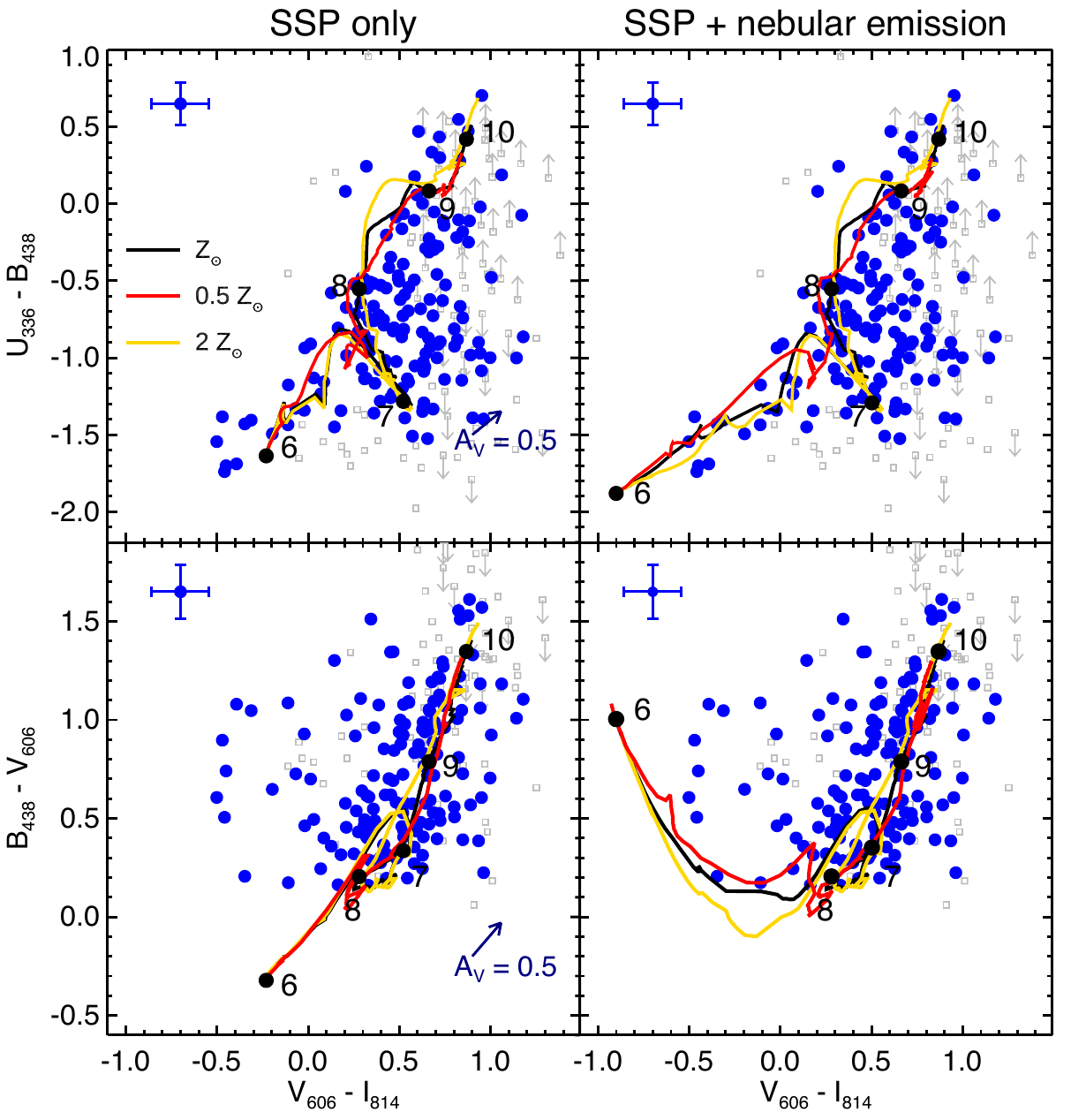}
		\caption{Comparison of SSP tracks at different metallicities, plotted against data for NGC 6872. Data points are objects 
			that fall within the tail, as in Figure \ref{fig:6872_all}. Black lines correspond to
			Z = Z$_\odot$, red corresponds to Z = 0.5 Z$_\odot$, and gold corresponds to Z = 2 Z$_\odot$. On the left, we plot tracks
			from \citealp{marigo_08}. On the right, we include nebular emission (see Section \ref{sec:nebs}) from Starburst99 models \citep{sb99};
			the inclusion of line emission affects only ages
			< 10 Myr. In both 
			cases, for a pure SSP and one with nebular emission added to it, the tracks are similar to one another; consequently, the ages and masses derived
			using either metallicity will be similar.}
		\label{fig:SSPs}
	\end{figure}

	Ages and masses of clusters were determined using the 3DEF spectral energy distribution (SED) fitting code, as described in \cite{bik_03}. This code compares a set of
	input magnitudes to a grid of SSP models with ages between 10$^6$ and 10$^{10.12}$ yr. It will apply an extinction to our observed magnitudes through
	a range of $E(B - V)$ values, compare the 
	set of model magnitudes and extincted, observed magnitudes, and minimize the resulting $\chi^2$ value to determine the best fit ages and masses.
	We use an evolutionary
	track from \cite{marigo_08} with solar metallicity and a Salpeter IMF \citep{salpeter_55}, with nebular flux added to it (see Section \ref{sec:nebs}). Our models
	do not account for binary star evolution.
	Spectroscopic observations of clusters and merging systems have found
	metallicities in the range of $\sim$ 1.0 - 1.5 Z$_\odot$ (\citealp{rosa_14,trancho_12,bastian_09,trancho_07_2,trancho_07_1}). For comparison,
	we run our 3DEF algorithm for NGC 6872, with 158 SCCs, using a tracks at 0.5 Z$_\odot$ and 2 Z$_\odot$. We find the median ratio of ages and masses
	determined using solar and half solar metallicity to be 1 and 1.1, respectively. The ratio of ages and masses between Z$_\odot$ and 2 Z$_\odot$ is
	1.02 and 0.98, respectively. A comparison of the different metallicities and their effects on our \U - \B, \B - \V, and \V - \I colours is shown in Figure \ref{fig:SSPs}.
	Here, we plot data for NGC 6872 against SSP models and SSP models with nebular emission added, for metallicities of Z$_\odot$, 0.5 Z$_\odot$, and 2 Z$_\odot$.
	The similarity between all three metallicities shows that our age and mass estimates will not be sensitive to our chosen metallicity.
	
	The choice of IMF will influence the masses of our clusters, but
	generally has little effect on the derived ages. A Chabrier or Kroupa IMF will decrease the masses by a factor
	of $\sim 2$ relative to our assumed Salpeter IMF.
	
\begin{figure*}
		\centering
		\includegraphics[width=0.8\linewidth]{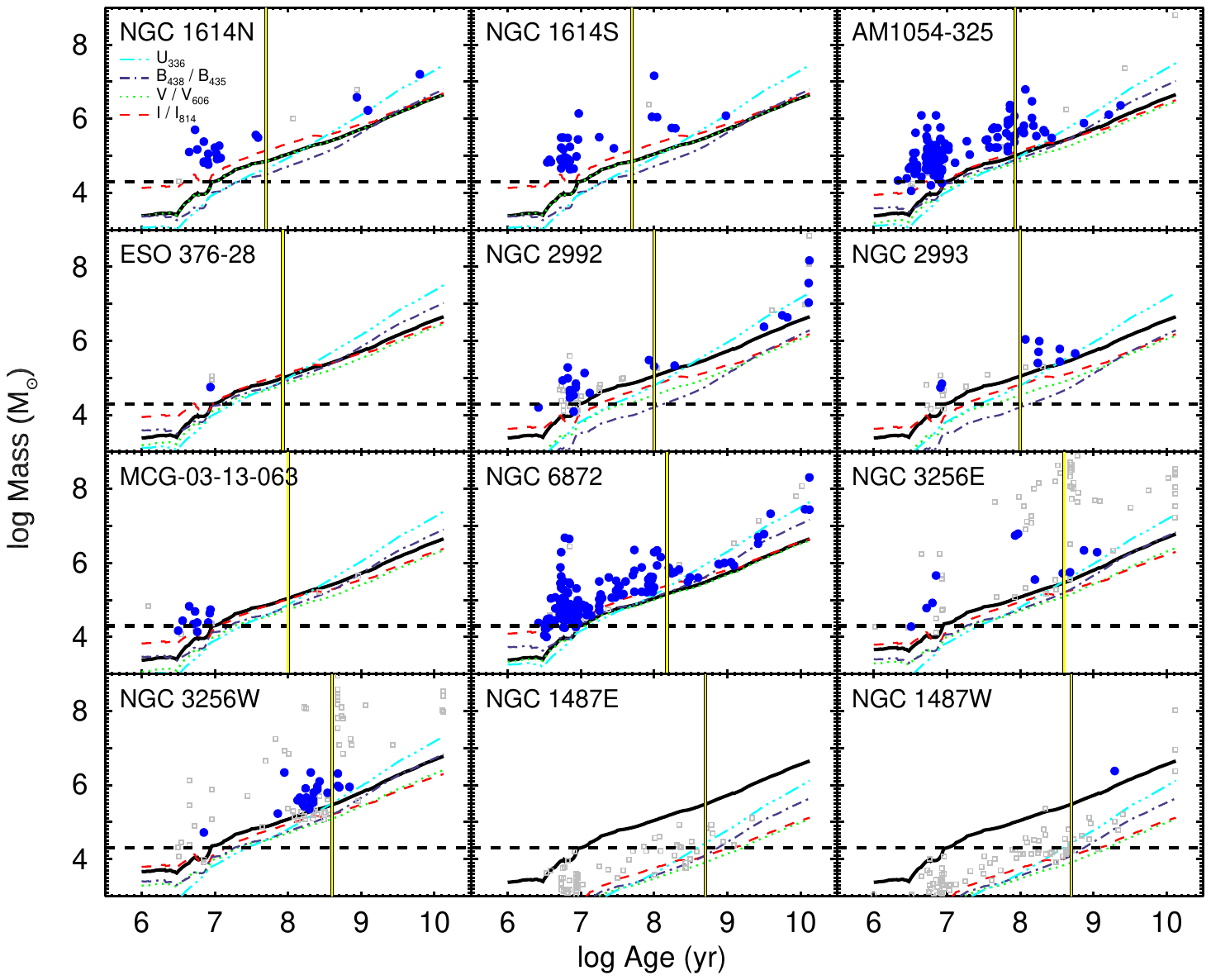}
		\vspace{-0.5cm}
		\caption{Ages and masses from Figures \ref{fig:1614N_all} - \ref{fig:1487W_all}, 
			compiled together. All systems except for NGC 1487 have at least one SCC with an age $<$ 10 Myr, 
			indicating recent cluster formation. Both NGC 3256W and NGC 3256E
			show a number of non-SCC objects at the interaction age, with large masses. Their high masses, indicating bright absolute magnitudes,
			along with poor fits to our SSP model ($\chi^2 > 3$, Section \ref{detect}) suggest these objects are foreground
			stars in the Milky Way. It is notable that these only appear in NGC 3256, which
			has the lowest Galactic latitude of our systems. They are not included in our SCC analysis.}
		\label{fig:mass_age_2}
	\end{figure*}

	Ages and masses for all our systems are plotted ndividually in Figures
	\ref{fig:1614N_all} - \ref{fig:1487W_all}, and collectively in Figure \ref{fig:mass_age_2}. We see a gap in age from log Age = 7.0 to 7.5 yr in many of our systems; this is an artifact of the fitting process and is seen in 
	similar studies of star clusters (e.g. \citealp{rand_19}; \citealp{de_grijs_13}; \citealp{bastian_05}). The true ages are likely spread over neighboring ages. 
	It is worth noting, however, that every system, with the exception of NGC 1487, contains SCCs with ages $<$ 10 Myr.
	This shows us that tidal debris are capable of supporting cluster formation.
	
	\subsection{Cluster radii} \label{radii}
	The radii of our SCCs were found using the program ISHAPE \citep{larsen_99}. ISHAPE requires that 
	sources are isolated to prevent PSF blending and at a high S/N. We perform an additional cut
	on our SCCs by visually selecting objects which are not in crowded regions and do not have highly elliptical shapes,
	which could indicate we are looking at blended clusters. We choose objects with a S/N $>$ 25 as determined by
	ISHAPE, as the program also	requires bright objects to perform accurate measurements.
	
	ISHAPE will deconvolve 
	selected sources with a user supplied PSF and a selected analytic model. Both an EFF \citep{eff_87} and 
	King \citep{king_62} model have been used in the past to study star clusters. We use a King model as it has been
	used to describe extragalactic and Galactic globular clusters \citep{correnti_21,larsen_21,chandar_16,bastian_12}. ISHAPE will fit the model to an ellipse and
	produce the FWHM of the major 
	axis, and the ratio between the minor and major axes (\textit{q}), for each selected
	source. We take the FWHM as the average between the major and minor axes of the
	fit. We require that \textit{q} > 0.3 to eliminate unrealistically elliptical sources, as in
	\cite{brown_21}.
	ISHAPE is
	able to reliably determine a FWHM for sources at 10\% of the size of the PSF, which for our WFC3 and ACS images,
	corresponds to $\approx 0.2$ pix (the pixel scale for WFC3 is 0.04'' per pixel, and 0.05'' per pixel for ACS). 
	We remove objects smaller than this size. The result of our selection criteria
	reduces the number of SCCs for analysis from 425 to 57, largely as a result of selecting isolated clusters. Our furthest
	systems, NGC 6872 in particular, are susceptible to crowding as the angular separation for nearby objects decreases,
	and this is borne out in our reduced number of sources. As a result, we emphasize that our radii sample
	is biased towards isolated, physically large SCCs.
	
	This FWHM value is converted to a half-light radius r$_h$ (also referred to as the effective radius R$_{eff}$) 
	by multiplying the FWHM by 1.48, as noted in the ISHAPE manual. We note that while an EFF and King model will produce unique FWHM values, the resulting half-light radii are similar
	to one another \citep{larsen_99}. The minimum value of r$_h$ we are able to detect, when combining our FWHM and axis ratio limits
	and converting FWHM to r$_h$, is 0.19 pix, corresponding to 0.0076'' for WFC3 and .0095'' for ACS. We use our \B- and
	\BA-band images to derive radii as they offer a better S/N than our \U-band images. 
	
\begin{figure*}
		\centering
		\includegraphics[width=0.7\linewidth]{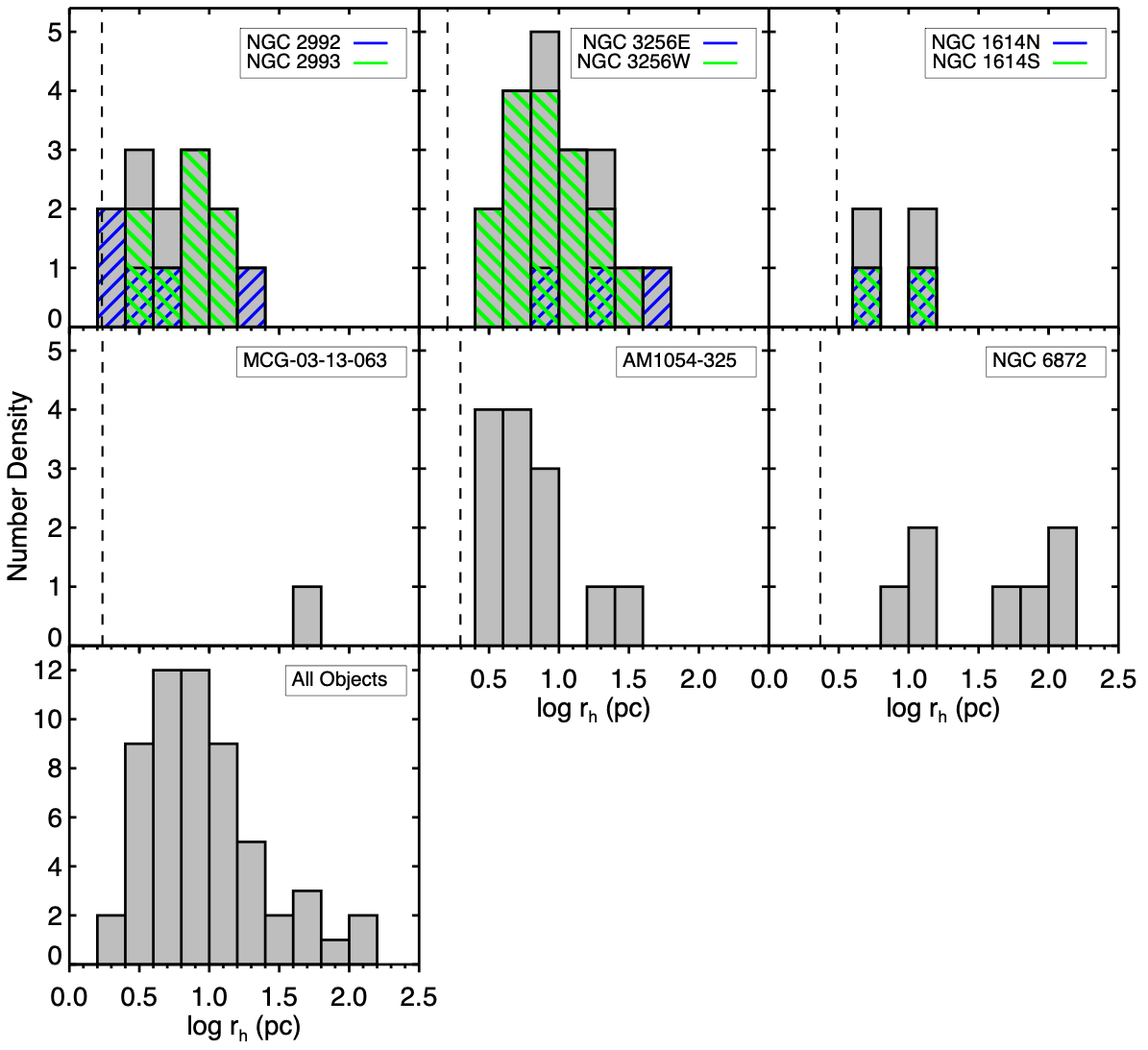}
		\caption{Half-light radii for our sources. Gray boxes indicate counts per bin for each 
			system as a whole, while coloured lines represent individual tails (where applicable). The vertical
			dashed line represents the minimum size of an object able to be detected by ISHAPE (0.2 pix) in an image.}
		\label{fig:all_radii}
	\end{figure*}

	Our cluster radii are shown in Figure \ref{fig:all_radii}, with interacting pairs grouped together. 
	The minimum detectable radius for each system is
	shown as a vertical dashed line. The combined distribution for all our sources is shown as the bottom left plot. The distribution
	shows a peak at $\approx 5.6$ pc, with an extended tail up to $> 100$ pc. Objects at the tail end of the distribution are likely
	blended together, but are included for completeness. 
	While our objects peak at a larger value than typical of Milky Way globular
	clusters ($\sim3.2$ pc, \citealp{baum_18}), objects of this size are seen in both the Milky Way and in extragalactic systems \citep{baum_18,ryon_17,chandar_16}.
	
	Note that ESO 376-28 is not included as it only contained
	one SCC, which was eliminated due to a low S/N in ISHAPE, and NGC 1487W's single SCC did not have a good fit. We address
	NGC 1487 in Section \ref{sec:1487}.
	
	\section{Results and Discussion} \label{sec:4_2}
	\subsection{Mass Function}
	
\begin{figure}
		\centering
		\includegraphics[width=1\linewidth]{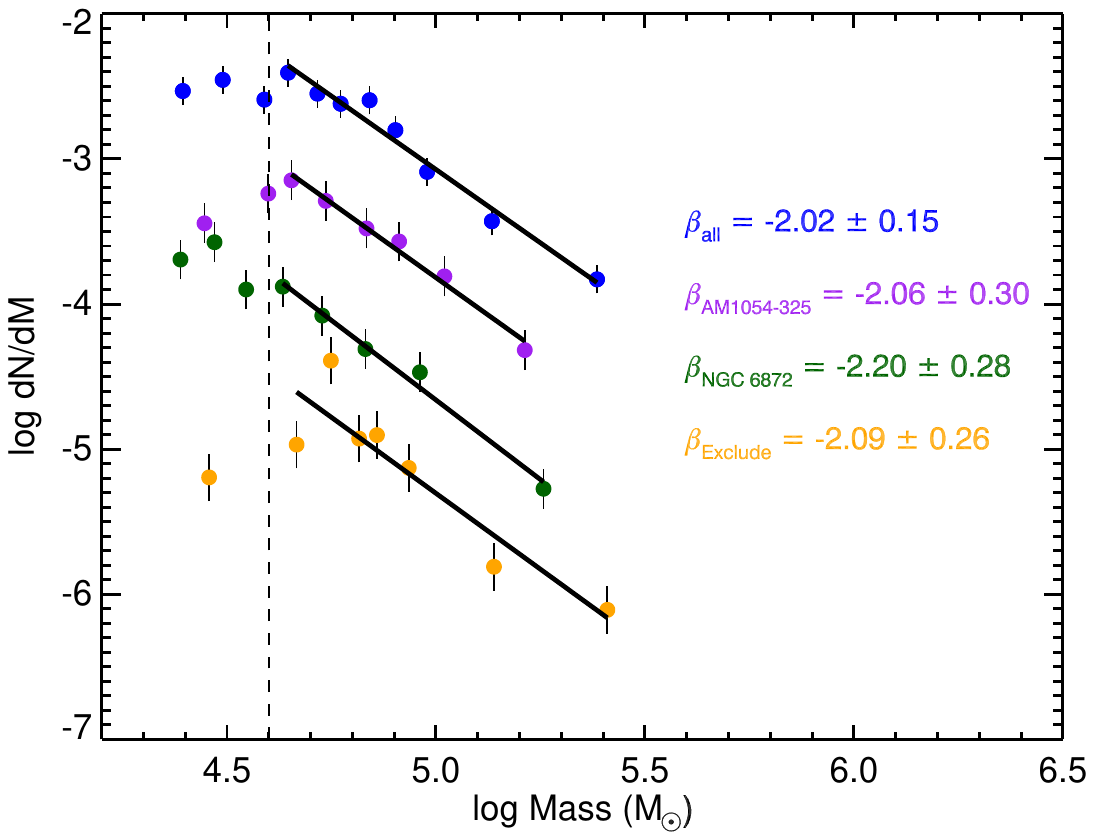}
		\vspace{-0.5cm}
		\caption{Mass functions for all, stacked objects (blue), AM1054-325 (purple), NGC 6872 (green), and 
			stacked (orange), but excluding AM1054-325 and NGC 6872. Data are normalized to the stacked mass function
			and vertically offset to include all curves on the same plot. The corresponding fit slopes
			are shown on the right. While there is some scatter in $\beta$, all values are consistent with
			each other, within their uncertainties. Vertical dashed line indicates our cut-off mass at log Mass = 4.6.
		}
		\label{fig:MF_combo}
	\end{figure}

	A global mass function for all our systems is presented in Figure \ref{fig:MF_combo}. We stack our measured masses together for all 
	our SCCs and plot them with bins of constant number, at 20 per bin, with 235 SCCs total. The mass bins are modeled as a power law with the form
	$dN/dM \propto M^{\beta}$. The slope, $\beta$, is found with a linear fit to log $(dN/dM)$ with $\beta = -2.02 \pm 0.15$. The lower mass cutoff for our fit
	is at log $M$ = 4.6 \msol, where the mass function begins to turnover. This turnover is caused by incompleteness at the lower mass limit.
	While some studies have suggested the mass function follows the form of a Schecter function \citep{messa_2018,adamo_15,bastian_12}, we see no turnover at high masses, leading
	us to conclude a power law is a good fit for our data.
	
	The value of $\beta$ has been measured for many other systems as well, with values ranging from $\approx$ -2.15 -- -1.85. Our result of $\beta =$ -2.02 $\pm 0.15$ for our stacked distribution is
	consistent with these previous results. Varying the number of objects per bin finds values of $\beta$
	consistent with our stated value, with $\beta = 2.14 \pm 0.11$ and $\beta = 2.08 \pm 0.19$
	for 10 and 30 objects per bin, respectively.
	
	The low numbers of SCCs in each system means we cannot
	generate a useful mass function for each system. It is, however, useful to look at our two systems
	with the largest numbers of SCCs, AM1054-325 and NGC 6872. These contain 89 and 87 SCCs below 10 Myr,
	respectively. As these constitute more than half of the objects in our stacked mass function, it is
	possible that they have a heavy influence on the derived slope and form of the distribution. To ensure
	we are not affected by this, we also look at the mass function while excluding objects from
	AM1054-325 and NGC 6872. In Figure \ref{fig:MF_combo} we include these three cases to compare to our full,
	stacked function.
	
	The peak of the function at log Mass = 4.6 M$_\odot$ is normalized to the stacked function, and vertically offset 
	in steps of 0.75 dex to plot all functions on the same plot. We again plot data using bins of constant
	number, 10 for AM1054-325 and NGC 6872, and 7 for the excluded function. All cases
	are consistent with one another, and the stacked mass function as well, suggesting our more populous systems are not over-influencing the
	stacked mass function.
	
\begin{figure}
		\centering
		\includegraphics[width=1\linewidth]{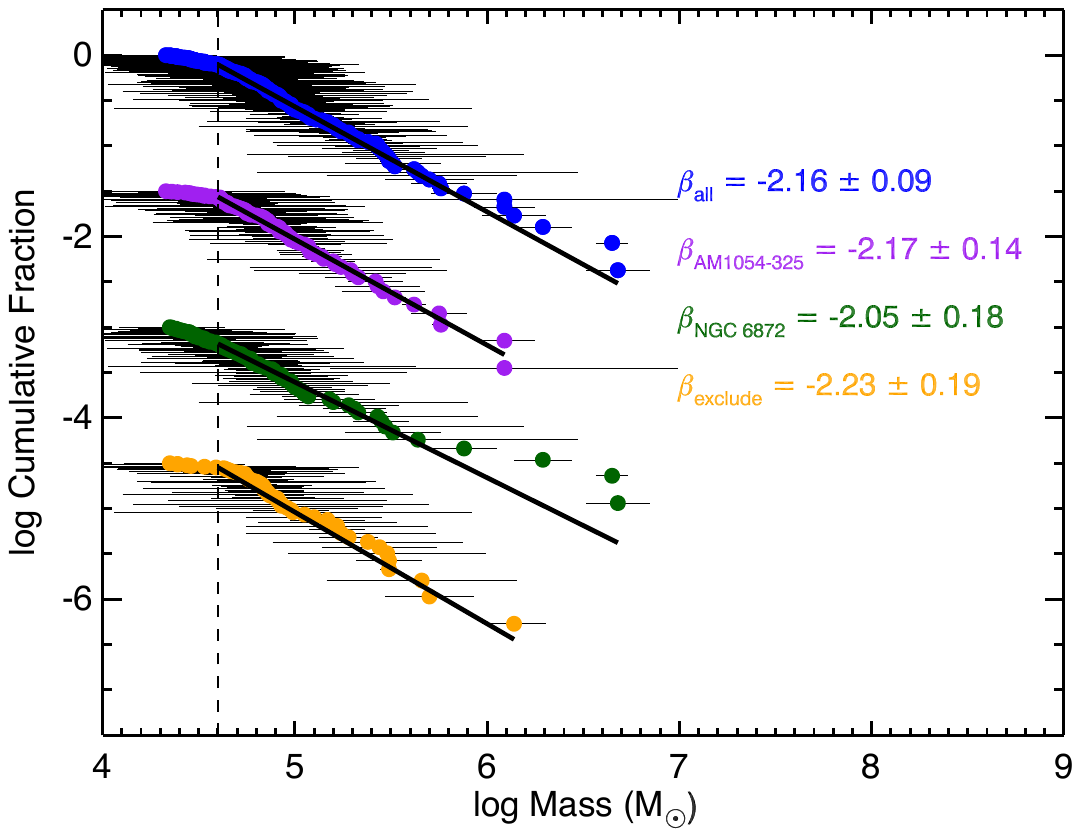}
		\vspace{-0.5cm}
		\caption{Cumulative mass distribution for stacked objects (blue), AM1054-325 (purple), NGC 6872 (green), and 
			stacked (orange), but excluding AM1054-325 and NGC 6872. Data are vertically offset to include all curves on the same plot. The corresponding fit slopes
			are shown on the right. Values of $\beta$ shown here are consistent with those from Figure \ref{fig:MF_combo} to within a standard deviation.
			Horizontal gray bars indicate 1$\sigma$ error bars for the individual mass points. Vertical dashed line indicates our cut-off mass at log Mass = 4.6.
		}
		\label{fig:MF_CDF}
	\end{figure}

	As a separate check, we fit our data to a power law using the the IDL program \texttt{mspecfit.pro}, developed
	by \citet{rosolowsky_05}. This uses a maximum-likelihood fitting technique to the cumulative distribution to
	find the slope of a power law distribution, and factors in the individual errors in mass for each data point. It will perform a fit to a regular power law $N(M' > M) \propto M^\beta$
	as well as
	determine the possibility of the cumulative distribution having the form
	of a truncated power law, given as $N(M' > M) \propto N_0[(M/M_0)^\beta - 1]$, where $M_0$ is the cutoff mass and $N_0$ is the number of objects
	more massive than $2^{1/\beta}M_0$. If $N_0 \lesssim 1$, then the distribution is best fit to a single power law. We fit the four 
	datasets shown in Figure \ref{fig:MF_combo} and find that $N_0 = 0.0 \pm 1.8, 1.0 \pm 2.0$,$0.0 \pm 0.8$, and $0.0 \pm 1.3$ for 
	our stacked mass function, AM1054-325, NGC 6872, and our excluded function, respectively. This suggests our data is best fit with a power law
	distribution. We thus plot our cumulative distribution functions of our data in Figure \ref{fig:MF_CDF}, along with the best fit 
	values of $\beta$, using a standard power law function. We find our results are consistent with that of our binned data within a standard deviation.
	
	\subsection{Cluster formation efficiency} \label{CFE}
	Stars form in clustered fashion, as seen in the Milky Way and in galaxies beyond \citep{bressert_10,whitmore_10,lada_03,clarke_00,zepf_99}.
	Observational studies have attempted to determine how many stars are born in bound clusters by measuring the amount of
	star formation occurring in clusters, compared to the local region (e.g. \citealp{adamo_15,ryon_14,goddard_10}). The cluster formation 
	efficiency (CFE) is defined as the ratio of the cluster formation rate and the star formation rate (both in units of
	\msol / yr), so that CFE = CFR / SFR. 
	Difficulty arises in the definition of ``bound'' clusters,
	as internal and environmental processes are capable
	of disrupting and unbinding clusters \citep{krumholz_19,zwart_10}, requiring age limits for cluster analysis.
	
	To match previous observations, we limit our SCC ages to 
	1 - 10 Myr. The SFR is measured from the UV flux of our
	\textit{GALEX} and \textit{Swift} images and converted to SFR using the following relation from \cite{kennicutt_12}:
	\begin{equation}
		\text{log SFR} \: (\text{M}_\odot \: \text{yr}^{-1}) = \log L_x - \log C_x,
	\end{equation}
	\noindent where $L_x$ is luminosity (ergs/s), and $C_x$ is a calibration constant dependent on the observed wavelength
	(43.35 for \textit{GALEX} FUV and 43.17 for \textit{Swift} \textit{uvm2}).
	
	The CFR was found by summing the mass of all clusters with an age between 1 - 10 Myr, and dividing by the time interval of 9 Myr.
	Completeness will affect the total mass of clusters which can be detected; low-mass clusters will not be observed in our images. We 
	find the approximate mass of a 10 Myr old cluster with M$_V$ = -8.5 (from our detection cutoff in Section \ref{detect}) to be $\approx$ 10$^{4.3}$ \msol.
	We assume we are complete above this mass limit, and calculate the missing, or undetected mass, assuming the cluster mass function
	follows a power law with slope $\beta$ = -2.0, extending from 100 \msol to 10$^7$ \msol. We perform the same calculation with slopes
	of $\beta$ = -2.15 and -1.85 for our upper and lower error bounds. For $\beta$ = -2.0, -2.15, and -1.85, we find the total percentage of mass in clusters with
	masses greater than 10$^{4.3}$ \msol\ to be 54\%, 33\%, and 74\%, respectively. Note that a shallower slope implies we measure a more complete sample of 
	our clusters.
	
\begin{figure}
		\centering
		\includegraphics[width=1\linewidth]{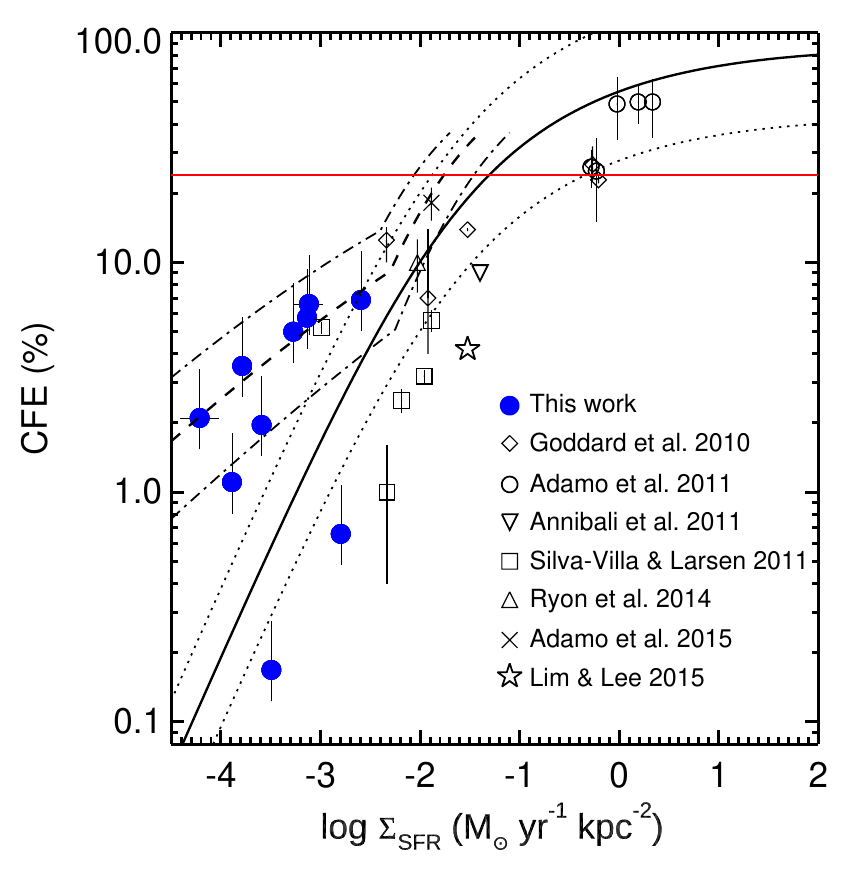}
		\vspace{-0.5cm}
		\caption{Cluster formation efficiency (CFE) plotted against the log star formation rate
			density. Included are several similar measurements gathered in \protect\cite{adamo_15}, as well as a
			theoretical curve from \protect\cite{kruij_12} in solid black. Dotted lines indicate a factor
			of 2 variation in the theoretical model. We include a modified version of the \protect\cite{kruij_12}
			curve, using the $\Sigma_{\text{gas}}$ vs $\Sigma_{\text{SFR}}$ relation from \protect\cite{bigiel_08}, as
			a dashed line. The dot-dash lines indicate upper and lower limits on this relation as defined by
			\protect\cite{johnson_16}. The red solid line corresponds to the 24\%
			value of CFE from \protect\cite{chandar_17}.}
		\label{fig:CFE}
	\end{figure}

	Our results are shown in Table \ref{table:CFE}. As NGC 1487W/E both do not contain clusters which fit our criterion, they do not have a 
	corresponding CFE. The general trend is that systems with a higher SFR density ($\Sigma_\text{SFR}$) show more efficient cluster formation. Data are plotted in Figure \ref{fig:CFE}. We plot our data against a theoretical CFE from 
	\cite{kruij_12} for comparison, with dotted lines indicating a variation factor of 2. This model
	predicts cluster formation to follow the surface gas density ($\Sigma_\text{gas}$), and therefore 
	$\Sigma_\text{SFR}$ via the Kennicutt-Schmidt law \citep{kennicutt_98}. The CFE tails off at high 
	$\Sigma_\text{SFR}$. At the high gas densities implied by the high $\Sigma_\text{SFR}$
	tidal interactions between neighboring GMCs can impede cluster formation. 
	
	While the Kennicutt-Schmidt law assumes the relation between $\Sigma_\text{gas}$ and $\Sigma_\text{SFR}$ 
	remains the same at all scales, this may not be the case. When studying star formation at sub-kpc scales,
	\cite{bigiel_08} found a decrease in $\Sigma_\text{SFR}$ at low values of $\Sigma_\text{gas}$. This decrease 
	is correlated
	with saturation of \HI\ in the total gas content of the region (\HI\ $+$ H$_2$). This effect has been parameterized 
	by \cite{johnson_16}, using a broken power law, to predict $\Sigma_\text{SFR}$ based on $\Sigma_\text{gas}$. We
	include the CFE as a function of $\Sigma_\text{SFR}$ based on this relation, again using the
	formulation from \cite{kruij_12}, in Figure \ref{fig:CFE} as a dashed line. The broken power law formula
	manifests as a flattening of the curve at -2.3 in log $\Sigma_{\text{SFR}}$, corresponding to the
	gas environment being dominated by \HI\ over H$_2$. Upper and lower limits
	provided by \cite{johnson_16}, derived from the \cite{bigiel_08} data, are plotted as dot-dash lines.
	
	We find a good agreement with the \cite{bigiel_08} curve, with only NGC 3256W and NGC 1614N falling more than
	1$\sigma$ below the lower limit. Notably, our data fall on the flattened part of the CFE curve, corresponding to
	an \HI\ dominated environment, suggesting the star forming regions in our tails are primarily composed of 
	\HI\ over H$_2$.
	
	We note that relation between the CFE and $\Sigma_\text{SFR}$ was designed for a ``typical'' spiral galaxy, 
	and should used as a rough estimate \citep{kruij_12}.
	In Figure \ref{fig:CFE}, we include data from previous surveys \citep{adamo_15,lim_15,ryon_14,adamo_11,annibali_11,silva_11,goddard_10} 
	as compiled by \cite{adamo_15}. Our
	data extend to smaller $\Sigma_\text{SFR}$ values than previously measured.
	
	The dependence of the CFE on $\Sigma_{\text{SFR}}$ has been questioned by \cite{chandar_17}, who studied
	several systems and found an average CFE of 24\%, independent of $\Sigma_{\text{SFR}}$. This constant value is 
	indicated in Figure \ref{fig:CFE} as a red horizontal line. They suggested that
	the variability seen in the CFE by other authors was caused by using different cluster age intervals for
	different galaxies in determination of the CFE. The result was a biased sampling of the CFE using a cluster age range of 0 - 10 Myr for systems with
	high $\Sigma_{\text{SFR}}$, and 10 - 100 Myr for systems with a low $\Sigma_{\text{SFR}}$.
	Indeed, we note that of the data points in Figure \ref{fig:CFE}, 
	\cite{adamo_15,lim_15,annibali_11,adamo_11} use objects with ages of 1 - 10 Myr, while the others use
	a range of ages from 1 - 10, 10 - 100, and 1 - 100 Myr in their CFE determinations. Our use of 1 - 10 Myr
	matches the majority of the listed studies. Independent
	of other studies, however, we find that our sample seems to match the theoretical model, and shows a trend
	of decreasing CFE with decreasing $\Sigma_{\text{SFR}}$.

	\begin{table*}
				\caption{SFRs and CFE for our sample. CFE is determined by comparing the mass of clusters with ages below 10 Myr
			to the SFR within their respective tidal tail. The SFR is found using \textit{GALEX} and \textit{Swift} UV
			data, converted to a SFR \citep{kennicutt_12}.}
		\begin{tabular}{llllll}
			System & SFR       & SFR$_{\text{density}}$ & CFE                                & A$_{\text{in}}$ &  $\Sigma_\text{SCC}$   \\ 
			& ($10^{-2}$ M$_\odot$ yr$^{-1}$) & ($10^{-3}$ M$_\odot$ yr$^{-1}$ kpc$^{-2}$) & (\%)                                      & (kpc$^2$)  & (kpc$^{-2}$)       \\ \hline \hline \vspace{1mm}
			ESO 376-28      & $0.55 \pm 0.25$  & $0.062 \pm 0.028$    & $2.1^{+1.3}_{-0.6}$ & 88.9  & $-0.015 \pm 0.013$ \\ \vspace{1mm}
			NGC 2993        & $2.31 \pm 0.15$   & $0.129 \pm 0.008$    & $1.11^{+0.70}_{-0.30}$     & 179.1  & $0.050 \pm 0.017$  \\ \vspace{1mm}
			NGC 2992        & $3.75 \pm 0.21$   & $0.163 \pm 0.009$    & $3.5^{+2.3}_{-1.0}$    & 229.5  & $0.070 \pm 0.022$  \\ \vspace{1mm}
			NGC 3256E       & $6.30 \pm 0.12$   & $0.256 \pm 0.005$    & $2.0^{+1.2}_{-0.5}$          & 246.0  & $0.034 \pm 0.014$  \\ \vspace{1mm}
			NGC 3256W       & $6.42 \pm 0.11$   & $0.321 \pm 0.006$    & $0.17^{+0.11}_{-0.05}$             & 199.9 & $0.103 \pm 0.025$  \\ \vspace{1mm}
			NGC 1487E       & $0.428 \pm 0.030$ & $0.345 \pm 0.024$    & N/A                & 12.4 & 0                  \\ \vspace{1mm}
			NGC 1487W       & $0.493 \pm 0.033$ & $0.360 \pm 0.024$    & N/A                                     & 13.7   & $0.073 \pm 0.073$  \\ \vspace{1mm}
			NGC 1614S       & $17.45 \pm 0.40$   & $0.532 \pm 0.012$    & $5.0^{+3.2}_{-1.4}$               & 327.8 & $0.078 \pm 0.019$  \\ \vspace{1mm}
			NGC 6872        & $65.4 \pm 3.0$   & $0.734 \pm 0.033$    & $5.7^{+3.7}_{-1.6}$                & 890.3  & $0.15 \pm 0.016$  \\ \vspace{1mm}
			MCG-03-13-063   & $0.98 \pm 0.33$    & $0.77 \pm 0.26$    & $6.6^{+4.2}_{-1.8}$             & 12.8     & $0.76\pm 0.25$  \\ \vspace{1mm}
			NGC 1614N       & $48.68\pm 0.14$   & $1.598\pm 0.005$    & $0.66^{+0.42}_{-0.18}$             & 304.6   & $0.046 \pm 0.017$  \\ \vspace{1mm}
			AM1054-325      & $33.0 \pm 3.1$    & $2.55 \pm 0.24$    & $6.9^{+4.4}_{-1.9}$            & 129.5    & $1.016 \pm 0.090$ \\ \hline
		\end{tabular}

		\label{table:CFE}
	\end{table*}
	
	\subsection{Spatial distribution}
	
	SCC age and mass are plotted against distance from the center of the interacting system in Figures \ref{fig:1614N_all} - \ref{fig:1487W_all}. The central point of each system is obtained as coordinates from SIMBAD. The median distance is marked with a dashed
	vertical line. We perform KS tests on the age and mass distributions
	for our SCCs as separated by the median distance to the center, with results shown in Table \ref{table:KS_dist}. We find statistically significant results beyond 2$\sigma$ in only two
	tails, NGC 1614S, and AM1054-325. NGC 1614S shows significant results in both its age and mass distributions, with \textit{p}-values
	of 0.0011 and 0.0052 for its age and mass distributions, respectively. From Figure \ref{fig:1614S_all} we see there is a clump of young objects
	near the base of the tail, with masses $\approx 10^5$ \msol. KS results for its companion tail, NGC 1614N, produce \textit{p}-values of 0.18 
	and 0.06 for the age and mass distributions, respectively. We do not claim either of these results for NGC 1614N to be significant.
	AM1054-325 shows significant results in the mass distribution, but not in age. 
	
	For the majority of our sample, we see that the general trend shows a relatively even distribution of ages though the tidal debris; young objects are not concentrated
	in any particular region, but are found throughout the tidal tails. 
	\begin{table}
				\caption{KS results for age and mass distributions of SCCs in tails, between objects interior
			and exterior to the median distance to the centre of the system.}
		\centering
		\begin{tabular}{lll}
			System        & KS$_\text{Age}$    & KS$_\text{Mass}$   \\ \hline \hline
			NGC 1614N     & 0.18   & 0.06   \\
			NGC 1614S     & 0.0011 & 0.0052 \\			
			AM1054-325    & 0.92   & 0.018 \\ 		
			ESO 376-28    & N/A     & N/A     \\
			NGC 2992      & 0.27   & 0.56   \\
			NGC 2993      & 0.96   & 0.96   \\
			MCG-03-13-063 & 0.44   & 0.99   \\
			NGC 6872      & 0.68   & 0.89   \\
			NGC 3256E     & 0.85   & 0.45   \\
			NGC 3256W     & 0.91   & 0.66   \\
			NGC 1487E     & N/A     & N/A     \\
			NGC 1487W     & N/A     & N/A     \\ \hline
		\end{tabular}

		\label{table:KS_dist}
		
	\end{table}

	\subsection{Dynamical age} \label{pi}
\begin{figure*}
		\centering
		\includegraphics[width=0.7\linewidth]{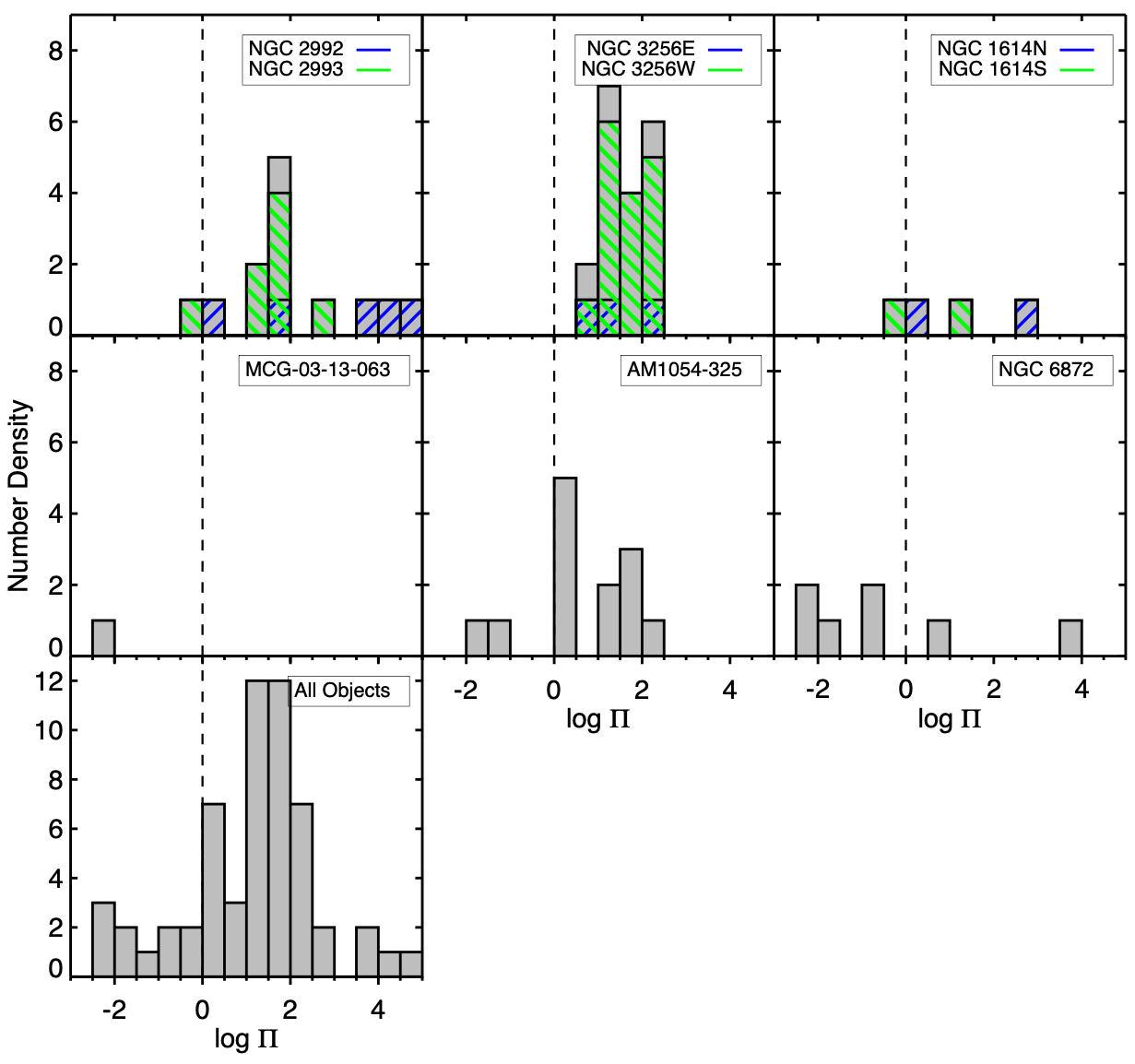}
		\caption{Dynamical ages of SCCs, following the prescription set by \protect\cite{gieles_11}. The vertical dashed line marks the limit for gravitationally bound objects; those to the
			right of the line are gravitationally bound, while those to the left are unbound. Gray boxes indicate 
			counts per bin for each 
			system as a whole, while coloured lines represent individual tails (where applicable).}
		\label{fig:radii_pi}
	\end{figure*}

	The dynamical age of a cluster ($\Pi$), introduced by \cite{gieles_11}, offers a method for 
	estimating 
	if a cluster is gravitationally bound at the current time. It is defined as the ratio of the age of a cluster
	to the crossing time of the cluster (\Tcr):
	
	\begin{equation}
		\Pi \equiv \frac{t_\text{cluster}}{T_\text{cr}}.
	\end{equation}
	$T_\text{cr}$ is defined as:
	\begin{equation}
		T_\text{cr}  \: (\text{s}) \equiv 10 \left( \frac{r_\text{h}^3}{G M} \right)^{1/2},
	\end{equation}
	
	where $G$ is the gravitational constant, $M$ is the mass of the cluster, and \rh\ is the half-light
	radius of the cluster. A cluster is said to be gravitationally bound if $\Pi \ge 1$. At this dynamical age, 
	the stars in the cluster have evolved for longer than 
	a crossing time, and as such are not likely to escape from the cluster, meaning the cluster is gravitationally bound.
	Objects with $\Pi < 1$ are not necessarily unbound, but as the stars have not yet evolved for longer than a crossing
	time, they still have time to escape before then, and we cannot determine if they are bound to the cluster.
	Unbound associations will expand with time, causing \Tcr\ to increase as well, and remain at or below $\Pi = 1$.
	Bound objects, on the other hand, will remain bound and compact with time, causing $\Pi$ to increase with time.
	Data for our SCCs are shown in Figure \ref{fig:radii_pi}.

	The determination of a cluster as bound or unbound remains an estimate, as several factors will influence the
	evolution of a cluster. If the natal gas of a cluster is expelled too quickly, on a timescale comparable to the 
	crossing time, the cluster is more susceptible to disruption, for a given star formation efficiency \citep{baum_07,hills_80}.
	Similarly, it is not possible for us to determine if a cluster, even with $\Pi > 1$, will remain bound
	throughout the lifetime of the merger, as this calculation does not take into account external forces. 
	Clusters which pass	nearby GMCs or through the disk or nucleus of the merging galaxies are subject to 
	gravitational tidal forces which can disrupt them (e.g. \citealp{krumholz_19,kim_18,tacconi_08,spitzer_58}). 
	However, as these objects exist in the diffuse regions of tidal debris, their chances of survival is
	increased as they will not experience the strong gravitational forces seen in the nuclear regions.
	
	\subsection{Age and mass as related to half-light radii}

	There is debate on the relation between cluster age and mass on their physical sizes. The initial half-light
	radius of a cluster is predicted, from an analytic model, to depend on the mass and gas surface density
	\citep{choksi_19}. As clusters evolve, mass loss and stellar interactions within the cluster will cause its
	expansion \citep{zwart_10}. \cite{gieles_11_2} suggest two stages of radii evolution for clusters in tidal fields: expansion, driven by mass loss,
	followed by contraction. They find that two-thirds of the Milky Way's globular clusters are in the expansion phase.
	
	Several studies of extragalactic clusters have found a slight dependence of age on radius \citep{ryon_17,bastian_12,lee_05_2}, while
	others have found none \citep{brown_21,scheepmaker_08,larsen_04}. There is a similar debate on the relation between
	mass and radius, and it is not clear that observational studies have found a relationship between these cluster
	properties \citep{brown_21,ryon_17}. There is considerable scatter in the distribution of cluster radius, hiding
	any clear signals.
	
\begin{figure}
		\centering
		\includegraphics[width=1\linewidth]{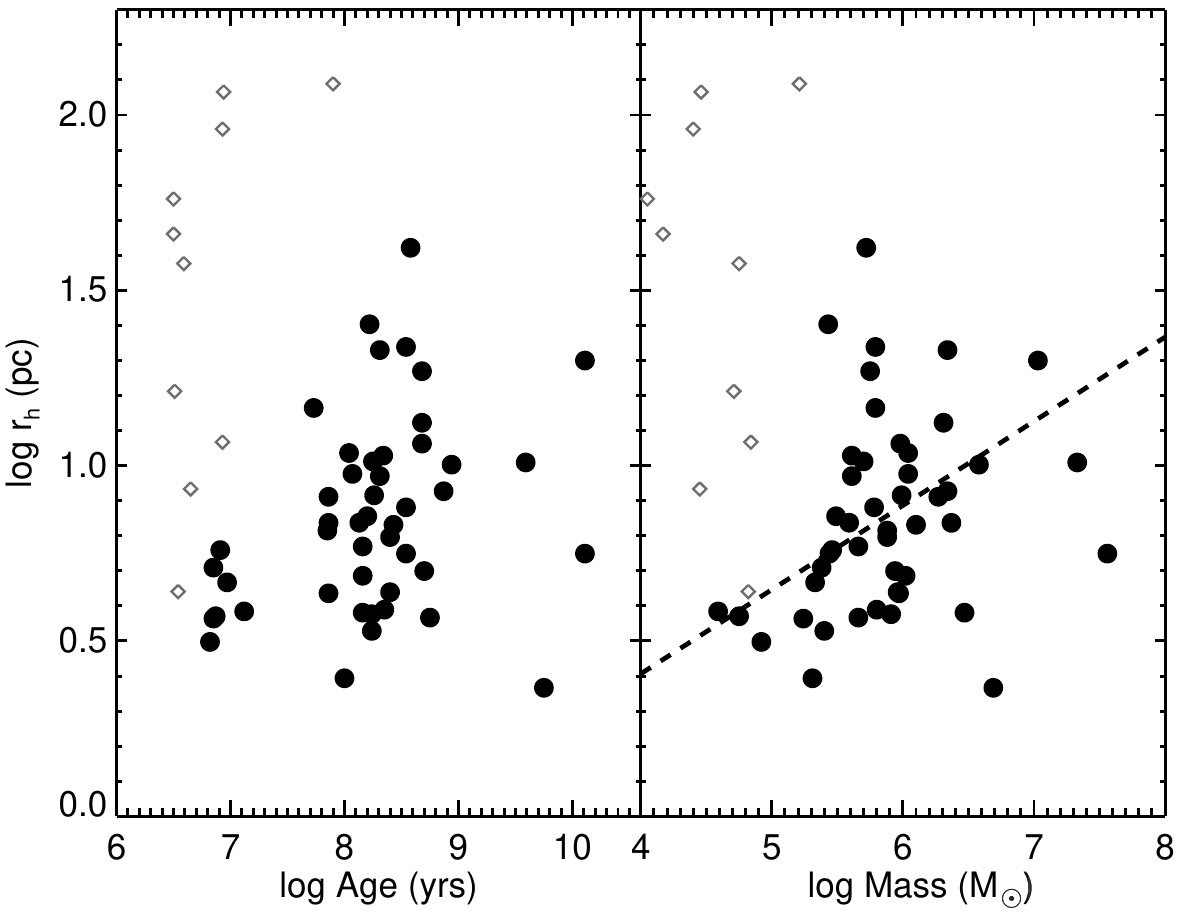}
		\vspace{-0.5cm}
		\caption{Half-light radii for our sources, as a function of age (\textit{Left}) and
		mass (\textit{Right}). Filled circles represent bound objects,
		while diamonds are unbound objects, using the definition set in Section \ref{pi}. On the right, we include
		the mass-radius relation from \protect\cite{brown_21}. Our data are consistent with their model, but there is a large
		amount of scatter in our data.}
		\label{fig:mass_age_radius}
	\end{figure}

	We plot the ages and masses of our SCCs against their radii in Figure \ref{fig:mass_age_radius}. We include
	both objects which are determined to be gravitationally bound and those which are not. We again find considerable
	scatter in our radii distribution, complicating any clear conclusions. On the left side of Figure
	\ref{fig:mass_age_radius} we compare the ages of our SCCs to their radii. There are only a small number of 
	objects at ages $< 10$ Myr, as a result of our selection criteria, namely that we look for isolated objects.
	Star clusters are formed in clustered fashion \citep{grasha_15,goul_15,bastian_09_2,gieles_08}, meaning
	our young SCCs will be in crowded regions which have been removed from our ISHAPE catalogue. This effect is seen
	in the fact that the young, unbound objects have large, extended radii, indicating blending. Thus, while 
	Figure \ref{fig:mass_age_radius} suggests that cluster radius increases with age, we have a small 
	and biased sample at
	ages around 10 Myr.

	Looking at the right panel of Figure \ref{fig:mass_age_radius} we see the relation between our masses and radii.
	We again see a large amount of scatter in our data, with no clear trend. Recent work by \cite{brown_21} looked
	at the cluster radii of 31 galaxies from the Legacy Extragalactic UV Survey (LEGUS). LEGUS galaxies are nearby
	($< 12$ Mpc) spiral and irregular galaxies, with few interacting systems and no major mergers. Their work 
	finds a power-law relation between mass and radius extending up to 10$^5$ \msol. Above this mass limit, the relation 
	appears to flatten, though this may be the result of low numbers of clusters with these large masses. We overplot
	their fit in Figure \ref{fig:mass_age_radius}, continuing their trend to higher masses. We find our data are 
	consistent with their fit, though there is large scatter.

	\subsection{NGC 1487 Objects} \label{sec:1487}
	
	

	NGC 1487 stands out from the rest of the mergers in our sample: it is three times closer than
	our next closest galaxy, there is only one SCC between the two tidal tails, and it has been classified
	as both a merger between two disk galaxies \citep{aguero_97} and a merger between dwarf galaxies
	\citep{buzzo_21,bergvall_03}. Despite the lack
	of SCCs, visual examination of the tails shows an abundance of objects within the debris. This suggests
	the debris host faint, low-mass objects which belong to the merging system, but are not luminous, high-mass 
	clusters with $M_V < -8.6$. The 
	absence of high-mass objects may arise if this is indeed a merger between dwarf galaxies.	
	Massive star clusters and high SFRs require high gas pressure \citep{maji_17,zubovas_14,blitz_06},
	and these required pressures may not be produced in a dwarf galaxy merger. \cite{lahen_19} were
	able to simulate a merger between two equal sized dwarf galaxies, which produced clusters with masses
	$\ge 10^5$ \msol. Pressures in these clusters was found to be $\sim 10^7$ $k$(K cm$^{-3}$)$^{-1}$, smaller
	than the $10^8 - 10^{12}$ $k$(K cm$^{-3}$)$^{-1}$ values seen in simulations of major mergers \citep{maji_17}.

\begin{figure}
		\centering
		\includegraphics[width=1\linewidth]{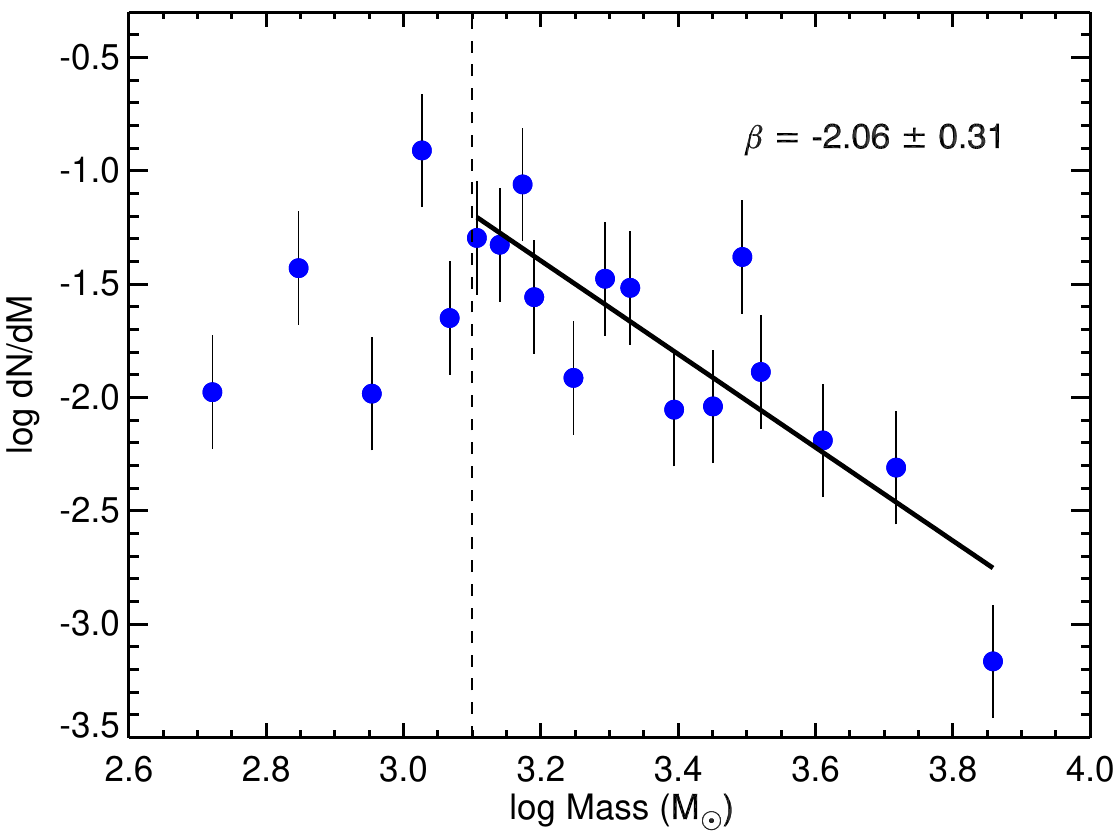}
		\vspace{-0.5cm}
		\caption{Mass function for sources in NGC 1487E and NGC 1487W. The dashed vertical line indicates our mass cut at log Mass = 3.1.}
		\label{fig:MF_1487}
	\end{figure}

	To consider	this scenario of low-mass cluster formation, we construct a stacked mass function with 
	objects from both tails. We eliminate our magnitude
	limit, but still require that sources are fit to our SSP models with $\chi^2 \le 3$ and \V\ - \I\ < 1.43. We
	only include objects with an age $\le$ 10 Myr as before; results are shown in Figure \ref{fig:MF_1487} for 58
	objects which meet our requirements. We fit our data
	to the mass turnover at log Mass = 3.1, and find that our data shows a slope of $\beta = -2.06 \pm 0.31$. This
	is consistent with our results at higher masses for our SCCs, where the completeness limit only
	allowed a fit down to log Mass = 4.6. 
	
	Our SED modelling assumes a continuously populated IMF, which is a reasonable assumption for
	our clusters with masses $> 10^4$ \msol. Below this mass limit, the stochastic sampling of the 
	stellar IMF can affect photometric measurements \citep{larsen_11}. The low numbers of stars can mean that a cluster
	can host only single digit numbers of supergiants, or none at all. The overall colour of the cluster
	would become bluer in the absence of supergiants, causing us to underestimate the age; consequently, the
	mass can decrease as well, as younger clusters are more luminous than older ones. This can result
	in a bi-modality of colours in clusters \citep{pop_10,silva_11}. Despite this, the effect on the slope of
	the mass function $\beta$ is small \citep{fou_12}, and eliminating sources with poor fits to models
	will reduce this effect \citep{fou_10}.
	
\begin{figure}
		\centering
		\includegraphics[width=1\linewidth]{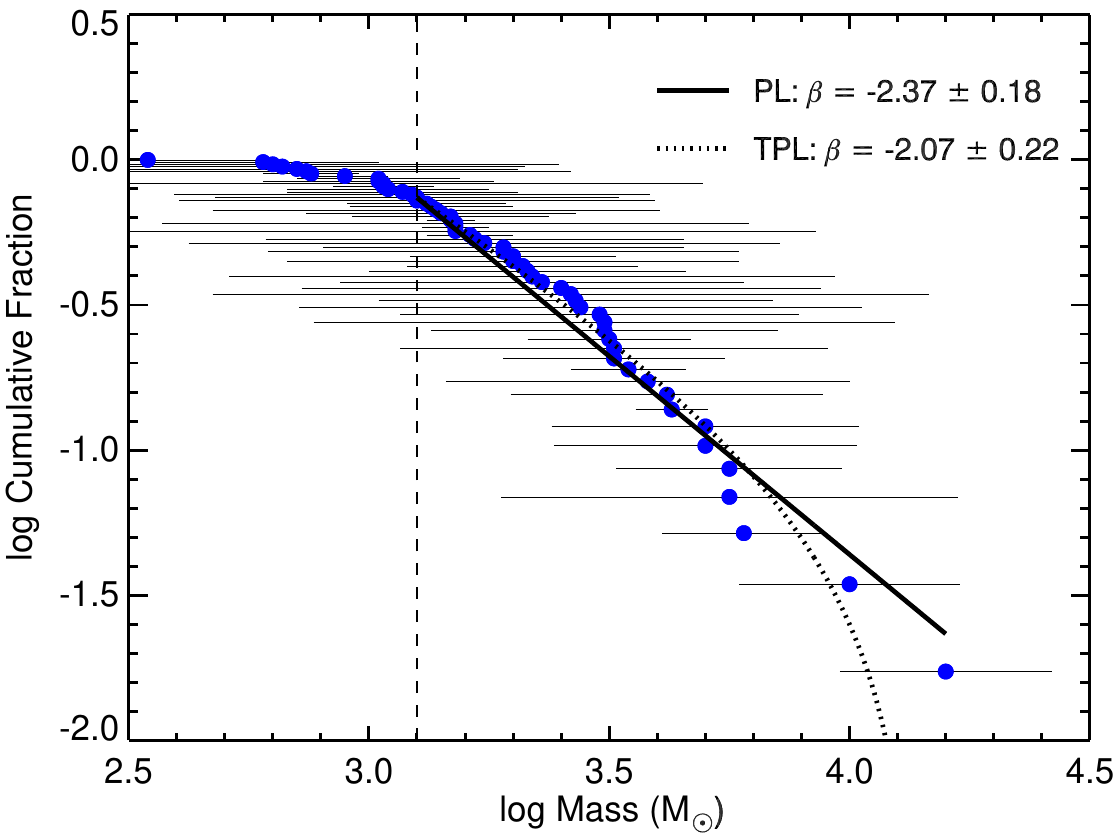}
		\vspace{-0.5cm}
		\caption{Cumulative fraction of sources in NGC 1487E and NGC 1487W. We overplot a power law function as a solid black line, and a truncated
			power law as a dotted line.}
		\label{fig:MF_1487_cdf}
	\end{figure}

	We plot the cumulative fraction of objects in Figure \ref{fig:MF_1487_cdf}, again using the \texttt{mspecfit.pro} code to 
	search for a truncated power law. We find for NGC 1487 a value of $N_0 = 3.7 \pm 3.6$, giving marginal
	significance ($\approx 1\sigma$) for a truncated power law. We plot both a standard power law and a 
	truncated power law to our cumulative mass distribution in Figure \ref{fig:MF_1487_cdf}.
	
	The similarities in slopes of our mass functions of NGC 1487 and stacked systems (see Figures \ref{fig:MF_combo} and \ref{fig:MF_1487}, and Figures \ref{fig:MF_CDF} and \ref{fig:MF_1487_cdf}) imply we are observing
	the low-mass end cluster formation.
	We suggest that the pressure in NGC 1487 is too low to reach the threshold for massive clusters, as it is the result of 
	mergers between dwarf galaxies, and not a major merger.
	
\begin{figure}
		\centering
		\includegraphics[width=1\linewidth]{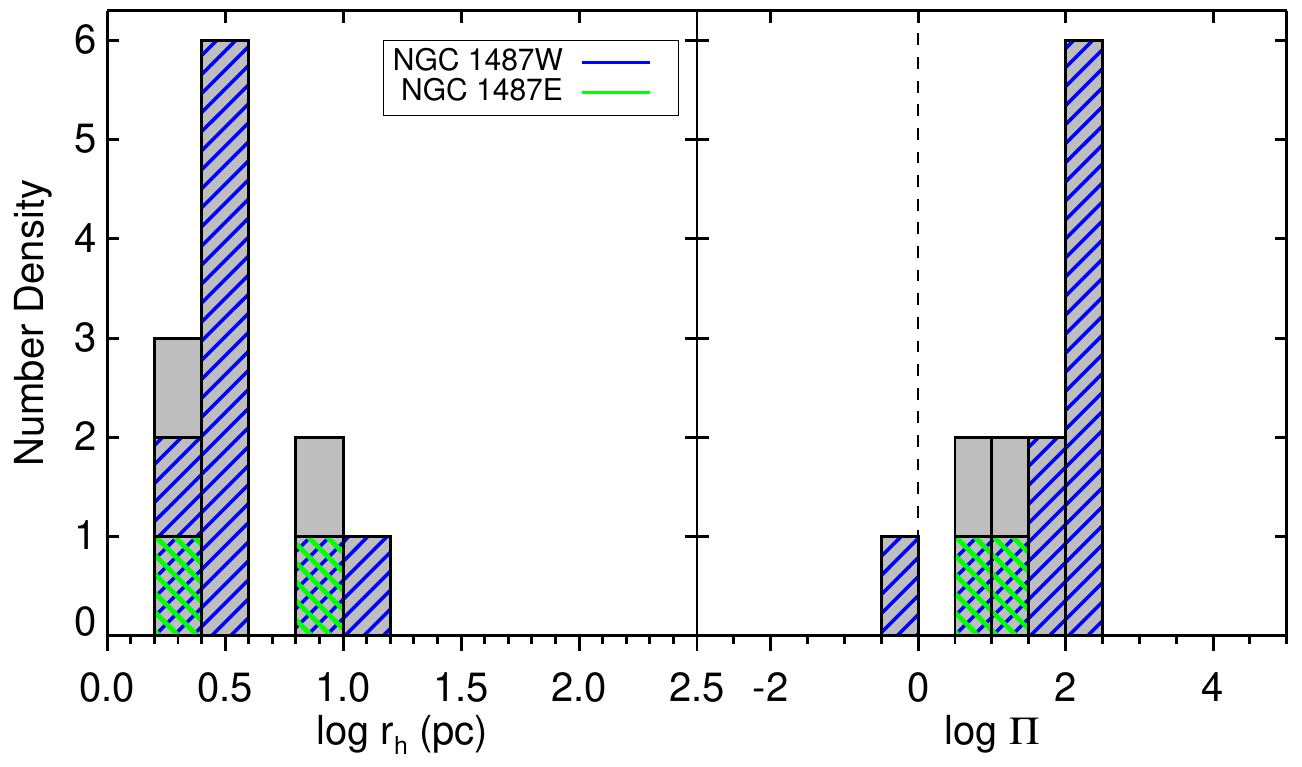}
		\vspace{-0.5cm}
		\caption{Half-light radii (\textit{left}) and dynamical ages (\textit{right}) for sources in NGC 1487W and NGC 1487E.}
		\label{fig:1487_r}
	\end{figure}

\begin{figure}
		\centering
		\includegraphics[width=1\linewidth]{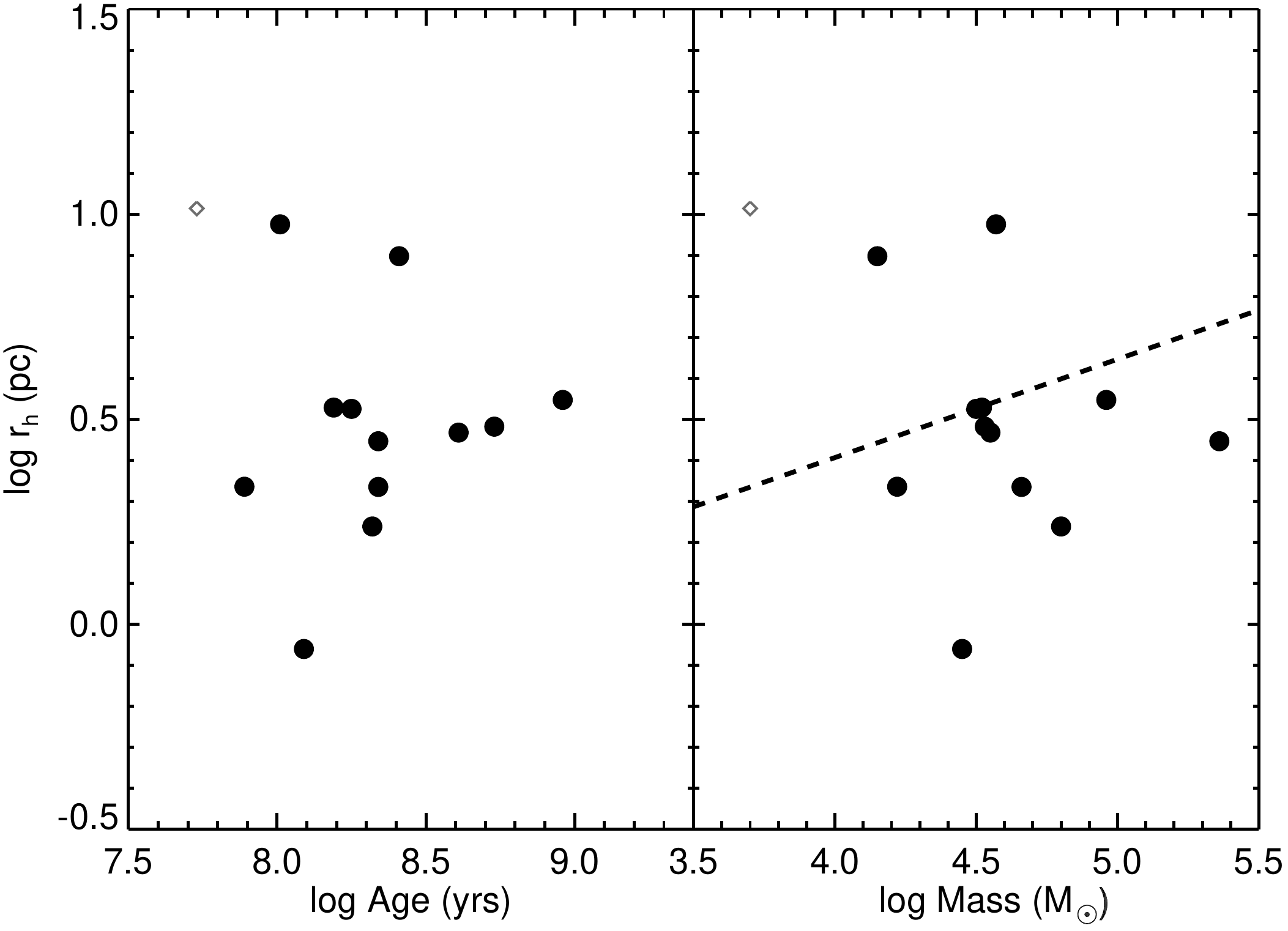}
		\vspace{-0.5cm}
		\caption{Same as Figure \ref{fig:mass_age_radius}, but for combined sources in NGC 1487E/W. We again
			include the mass-radius relation from \protect\cite{brown_21}.}
		\label{fig:1487_mass_age_radius}
	\end{figure}

	
	We look at the cluster radii of these objects as well, using ISHAPE.
	Neither tail in NGC 1487
	contains enough foreground stars to produce a PSF image; we generate a PSF for each of the five \B -band images in our sample (NGC 3256W, NGC 3256E, NGC 6872, AM1054-325, and MCG-03-13-063), run ISHAPE five separate times, using each
	PSF once. Our final derived \rh\ value is the mean value from each run. We find the dynamical ages for these objects
	as well, as in Section \ref{pi}. Results are shown in Figure \ref{fig:1487_r}. All objects except for one in these
	tails are gravitationally bound, following \cite{zwart_10}. Figure \ref{fig:1487_mass_age_radius}
	shows the age and mass of these objects plotted against their radii. We again include the relation
	of \cite{brown_21} between mass and radius. Our data points show similar scatter as our SCCs,
	in Figure \ref{fig:mass_age_radius}.
	
	Objects in NGC 1487 are more compact than in our other tails. The median half-light radius for bound sources
	in NGC 1487 is 3.03 pc, compared to 6.78 pc for all our other systems. We do not see the large,
	10 pc clusters that exist in NGC 3256 or NGC 2992/3.

	\section{Summary} \label{sec:5_2}
	
	We have analyzed 425 SCCs in 12 tidal tails, across seven merging systems. We summarize our findings as follows:
	
	\begin{enumerate}
		\item Many objects in tidal tails show signs of line emission in colour-colour diagrams, indicating young ages $< 10$ Myr.
		Clusters at these ages will have strong emission in H$\alpha$, which falls in our \V-band filter. The effect
		of this emission impacts our colour-colour diagrams. These colours indicate current, ongoing star formation
		in tidal debris. The age and mass distributions (Figure \ref{fig:mass_age_2}) suggest that previous
		star formation episodes produced many more SCCs, as evidenced by high-mass objects seen at older ages.
		
		\item The mass function of our SCCs has a consistent shape as compared to YMCs in other systems. Conventionally,
		the mass function of YMCs takes the form of a power-law with a slope of $\beta \approx -2.0$. Other studies
		have found evidence of a high-mass cut-off, suggesting the mass function follows a Schecter function instead.
		We do not find evidence for such a mass cut-off, and find our data fit well to a power law with slope
		$\beta = -2.02 \pm 0.15$ using binned data, and $\beta = -2.16 \pm 0.09$ for a cumulative 
		distribution fit.
		
		\item The CFE in tidal tails increases as the SFR density increases.
		We use \textit{GALEX} and \textit{Swift} UV imaging to determine the SFR in our tails. When compared to
		previous observations and theoretical predictions using a reformulation of the Kennicutt-Schmidt law from
		\cite{bigiel_08} and \cite{johnson_16}, we find 
		good agreement, implying the gas in the tidal debris is primarily \HI\ and that the CFE depends on the local environment. 
		Our data pushed this link to lower SFR
		densities than previously observed for cluster formation.
		
		\item Little dependence on galactic radii is seen for ages or masses of SCCs. Our KS tests reveal only NGC 1614S
		has significant differences in age or mass distributions with regard to galactocentric distance, while
		AM1054-325 shows significance in mass and distance only. Our other systems show that young clusters
		are distributed throughout the tails.
		
		\item Cluster radii of gravitationally bound objects, as determined using calculations from 
		\cite{zwart_10}, fall in the range of 2 -- 32 pc, with a median value of 7 pc. 
		We do not see a relation between age and radius, or mass and radius. Work by \cite{brown_21}
		suggests cluster radius increases with mass; we include their power-law determination, which
		is consistent with our data.
		
		\item Low-mass objects in NGC 1487, which fall below our magnitude limit of M$_V$ < -8.5,
		show a mass function with a slope of $\beta = -2.00 \pm 0.28$ using binned data. We
		find minimal significance ($\approx 1\sigma$) for a truncated power law with slope $\beta = 
		-2.07 \pm 0.22$, and a slope of $\beta = -2.37 \pm 0.18$ for a pure power law,
		using a cumulative distribution fit. Though the uncertainties are large, these values are consistent
		with our stacked SCC mass function, suggesting cluster formation is consistent down to low masses.
	\end{enumerate}

	\section{Acknowledgments}
	
	We would like to thank the anonymous referee for helpful comments 
	which have improved the quality and content of this paper.
	This research is based on observations made with the NASA/ESA Hubble Space Telescope obtained from 
	the Space Telescope Science Institute, which is operated by the Association of Universities for 
	Research in Astronomy, Inc., under NASA contract NAS 5–26555. These observations are associated with programs
	GO-7466, GO-10592, GO-11134, GO-14066, GO-14937, and GO-15083. Support for this work was provided by grant
	HST-GO-14937.002-A and HST-GO-15083.001-A. The Digitized Sky Survey was produced at the Space 
	Telescope Science Institute under U.S. Government grant NAG W–2166. The images of these surveys are 
	based on photographic data obtained using the Oschin Schmidt Telescope on Palomar Mountain and the 
	UK Schmidt Telescope. The plates were processed into the present compressed digital form with the 
	permission of these institutions. This research is based on observations made with \textit{GALEX}, obtained 
	from the MAST data archive at the Space Telescope Science Institute, which is operated by the Association 
	of Universities for Research in Astronomy, Inc., under NASA contract NAS 5–26555.
	
	\section{Data Availability}
	
	 \textit{HST}, \textit{GALEX}, and \textit{Swift} data are publicly available through the 
	 MAST portal at \url{https://mast.stsci.edu/portal/Mashup/Clients/Mast/Portal.html}. 
	 The derived data generated in this research will be shared on reasonable request to the corresponding author.
	
	
	
	
	\bibliographystyle{mnras}
	\bibliography{refs} 

	


\appendix

\section{Notes on Individual Tails} \label{sec:6_2}

\subsection{NGC 1614N/S}
This is our furthest merger in our sample, at a distance of 65.6 Mpc. The tidal debris are the youngest of our
		systems, forming 50 Myr after the second passage of the two galaxies \citep{vaisanen_12}. It boasts two tidal tails,
with the northern tail wrapping back onto the centre of the merger, and the southern tail extended radially outward.
This system has a large IR luminosity, with L$_{IR} = 4 \times 10^{11}$ L$_\odot$, classifying it as a Luminous Infrared Galaxy (LIRG).
The large IR luminosity appears to be driven through star formation. A 300 pc ring surrounding the nucleus shows evidence
of luminous H\textsc{ii} regions as revealed in Pa-$\alpha$ observations. This ring also coincides with 3 micron PAH emission 
and radio continuum \citep{vaisanen_12}.

X-ray observations initially suggested the presence of an obscured AGN \citep{Risaliti_00}. However, \textit{L}-band
observations of the nucleus do not show any enhanced emission \citep{vaisanen_12} as is characteristic of an AGN \citep{sani_08}.
The X-ray emission can be explained through low-mass x-ray binaries (LMXBs) \cite{Olsson_10}. Most notably, high resolution VLBI observations
do not find a compact source in the nucleus \citep{herrero-illana_17}. Should an AGN exist in NGC 1614, it is faint and turned off.

Figures \ref{fig:1614N_all} and \ref{fig:1614S_all} show the colour-colour diagrams of the SCCs in each tail. At a glance, the two tails in the same system appear different
from each other. The Southern tail has several objects with blue colours at \V\ - \I\ $< 0$, which do not appear in the Northern tail.
This is reflected somewhat in the age distributions for the two tails, as the Southern tails has several more young ($< 10$ Myr)
SCCs than seen in the Northern tail. However, the difference in \V\ - \I\ colour can be explained if internal extinction is affecting the Northern clusters. 

A KS test between the two tails for their SCC \U\ - \BA, \BA\ - \V, and \V\ - \I\ colours gives \textit{p}-values of 0.71, 1.3 $\times 10^{-5}$, and 5.2 x $10^{-9}$,
respectively. A KS test between their ages and masses gives \textit{p}-values of 0.049 and 0.69, respectively.

KS test results for NGC 1614S between in-tail and out-of-tail SCCs for \U\ - \BA, \BA\ - \V, and \V\ - \I\ colours give
$p$- values of 0.65, 0.045, and $1.2 \times 10^{-4}$, respectively. For NGC 1614N, we find $p$- values of
0.17, 0.19, and $7.2 \times 10^{-3}$, for \U\ - \BA, \BA\ - \V, and \V\ - \I. Cluster excess for NGC 1614S
and NGC 1614N are 0.078 $\pm 0.019$ kpc$^{-2}$, and 0.046 $\pm 0.017$ kpc$^{-2}$, respectively.
NGC 1614S has a cluster excess above 3$\sigma$, though NGC 1614N does not.

\subsection{AM1054-325 and ESO 376-28}

AM1054-325 and ESO 376-28 are an interacting pair in a relatively early stage of their merger history, with
an interaction age of 85 Myr. AM1054-325 is a spiral galaxy with an elongated spiral arm extending to the North and a tidal dwarf 
galaxy (TDG) candidate to the NE \citep{weil_03}. Spectroscopic measurements find H\textsc{ii} regions in AM1054-325,
created by shocked gas \citep{krabbe_14}, including some within the tidal dwarf galaxy, as well
as in the outer disk and extended spiral arm. Its interacting companion, ESO 376-28, is a lenticular galaxy. \cite{krabbe_14} finds
emission lines indicative of H\textsc{ii} regions within this galaxy as well.

ESO 376-28 contains only one SCC, with an age of 10$^{6.5}$ yr and a mass of 10$^{4.75}$ \msol, about 3.6 kpc from the
centre of the galaxy. Unfortunately, we cannot
derive a radius due to its low S/N in ISHAPE.

While ESO 376-28 is devoid of SCCs, its companion,
AM1054-325, hosts the largest excess of SCCs in our sample, at 1.016 $\pm$ 0.090 kpc$^{-2}$. A KS test of colours for in-tail and out-of-tail
SCCs gives $p$-values of 0.166, 4.5 $\times 10^{-3}$, and 2.0 $\times 10^{-8}$, for \U\ - \B, \B\ - \V, and \V\ - \I\, respectively.
We find $p$-values less than 0.04 for our \B\ - \V\ and \V\ - \I\ colours, indicating our in-tail and out-of-tail SCCs are drawn from differing distributions.
This is largely due to the young ages of the in-tail SCCs, as seen in the colour-colour diagram in Figure 
\ref{fig:AM1054_all}. Many objects have evidence of emission from recombination lines of hydrogen, notably H$\alpha$, strongly influencing the \V-band magnitudes. The
spatial distribution in Figure \ref{fig:AM1054_all} shows that the young, < 10 Myr SCCs are distributed through the length of
the tidal tail. Mixed in with the young SCCs, we see a group of older and more massive objects, with ages between 10$^8 - 10^{8.5}$
and masses between $10^{5.5} - 10^6$ \msol. These appear clustered around the TDG candidate, with ages similar to the age
of the interaction of the system; this region may have seen the first burst of star formation from the encounter.

From Figure \ref{fig:AM1054_all}, there appear to be two groupings of ages, one at 6.7 and one at 7.9. 
It is tempting to interpret this as two isolated bursts of star formation; however, 
it is possibly an indication of continuous star formation since the encounter, and we are 
instead seeing artificial effects due to age fitting. Similar artifacts
		are seen in other cluster studies (see Section \ref{age_mass}).	

The system is at a distance of 52.9 Mpc and is compact, with each galaxy filling one quadrant of the WFC3 FOV.
This causes crowding to be a concern when fitting radii, and we eliminated about a third of our SCCs. SCCs which
were fit show radii between $\sim 3 - 10$ pc. All but one of the 13 sources are gravitationally bound. 

\subsection{NGC 2992/3}
NGC 2992 and 2993 are disk galaxies in the initial stages of merging. This interaction has produced a tidal tail
in NGC 2993 and a tidal dwarf galaxy at the northern tip of NGC 2992. A tidal bridge, connecting the two main galaxies,
is not imaged in our data. Our \textit{HST} observations fall on the tidal tail of NGC 2993, and the TDG and associated tidal
debris, North of the main body of NGC 2992.

NGC 2992 contains an AGN and is classified as a Seyfert 1.9. The classification has
changed over time, as the AGN has shown variability, though observations from 2007 -- 2014 indicate it may have become stable.
The luminosity in the nuclear region appears to be dominated by the AGN, with low contributions from 
star formation. Dynamical models from \cite{duc_00} place the interaction at 100 Myr.

This system was observed in parallel observing mode, with ACS \BA\ and WFC3 \U\ imaging. Due to the observing
constraints, our \U-band imaging is not as deep as our other systems (see Section \ref{complete}) and has several more
undetected objects in \U, resulting in upper limits in Figures \ref{fig:2992_all} and \ref{fig:2993_all}.

A KS test between the SCCs for the two tails for their \U\ - \B, \B\ - \V, and \V\ - \I\ colours gives \textit{p}-values of 0.011, 0.16, and 1.2 $\times 10^{-3}$,
respectively. A KS test between their ages and masses gives \textit{p}-values of 0.40 and 0.47, respectively. This indicates the SCCs in both tails are consistent with being
drawn from the same distribution. The difference in KS results between colours and masses/ages can
be explained if the SCCs have internal extinction, resulting in different colours, but similar masses
and ages.

A KS test between SCC colours for in-tail and out-of-tail SCCs in NGC 2992 finds $p$-values of 0.73, 0.87, and 0.50 for \U\ - \BA, \BA\ - \V, and \V\ - \I\,
respectively. The fact that none of these values are below 0.04 means we cannot say that SCCs are drawn from
different populations. However, the SCC excess is
$0.070 \pm 0.022$ kpc$^{-2}$, above 3$\sigma$. Thus, while the colours of
in-tail and out-of-tail SCCs are similar, the strong excess of objects inside the tidal debris lends us 
confidence that these are not containment sources. NGC 2993 shows no SCCs out-of-tail, with
9 in-tail SCCs. While there are no out-of-tail SCCs to compare to, visual examination of Figure \ref{fig:2993_all}
shows an absence of detected objects (SCCs and non-SCCs) in the out-of-tail region, above \U\ - \B = -0.5,
suggesting objects in NGC 2993 are real.

The median SCC radius for the two tails is 5.6 pc, with all but one of our objects being bound. At the tip
of the tail in NGC 2993 we find a massive, extended cluster, with age of 10$^{8.1}$ yr, mass of 10$^{6.0}$ \msol, and a radius
of 9.5 pc.

\subsection{MCG-03-13-063}

MCG-03-13-063 is a disturbed spiral with no optical companion. Little data exists on this system. It features an extended spiral arm, hosting beads of clusters. The SCCs are recently formed, with ages $< 10$ Myr. The arm is thin, with clusters
located in well-defined regions \cite{mullan_11}.

Our KS tests for \U\ - \B, \B\ - \V, and \V\ - \I, find $p$-values of 2.1 $\times 10^{-3}$, 0.35, and 
1.7 $\times 10^{-3}$, respectively, for in-tail and out-of-tail SCCs.
The SCC excess in this system is $0.76 \pm 0.25$ kpc$^{-2}$, significant beyond 3$\sigma$.

\subsection{NGC 6872}
NGC 6872 is a disturbed barred spiral galaxy merging with the nearby, smaller galaxy IC 4970. It is located in the 
Pavo Group with 12 other members. The 
interaction has produced two elongated spiral arms which span a distance of 160 kpc from 
	East to West \citep{Eufrasio_14}.
Our study here focuses only on the Eastern tidal arm of the system; we do not image the Western
		tail. An X-ray view of the 
Pavo Group found a trail of x-ray emission between NGC 6872 and the large, dominant elliptical galaxy NGC 6876. The origin
of this trail is unlikely to be related to tidal interactions, as the morphology of NGC 6876 does not show any disturbances.
Furthermore, no \HI\ is seen associated with the x-ray emission \citep{Horellou_07}. Rather,
it is likely due to a Bondi-Hoyle wake; as NGC 6872 passes through the IGM, gas is gravitationally focused in its wake, and is compressed
and heated \citep{Horellou_07,Machacek_05}.

NGC 6872 has been modeled by \cite{mihos_93} and \cite{Horellou_07}, finding that the tidal arms can be reproduced in an interaction with a
small companion, at a mass ratio of 5:1. While models from \cite{mihos_93} predict enhanced star formation in the nucleus
of the galaxy, H$\alpha$ observations show a lack of star formation in the interior; rather, the star forming regions are
located in the tidal arms.

An isolated blue clump can be seen at the Northern tip of the Eastern tail in \textit{GALEX} imaging, which unfortunately falls outside
our FOV. This clump is a possible TDG, though spectroscopic analysis is needed to confirm that it is gravitationally 
bound \citep{Eufrasio_14}.

NGC 6872 hosts the largest numbers of SCCs in our sample, with 158 in the tail, and an excess of $0.150 \pm 0.016$ kpc$^{-2}$.
KS tests between in-tail and out-of-tail SCCs find $p$-values of 0.026, 0.049, and 0.025 for \U\ - \B, \B\ - \V, and \V\ - \I, respectively. 
Young SCCs are found throughout the tail,
from the base to the tip, suggesting this system is currently undergoing 
a burst of star formation. Our KS tests show that there is no statistical difference between ages or masses of
SCCs with distance. However, we see a visual trend for SCCs with ages > 10$^8$, where older SCCs are 
preferentially within our 41 kpc median distance. To a lesser extent, this extends to masses > $10^6$, with
more massive clusters located closer to the centre. Radii are difficult to determine as with NGC 1614,
due to the large distance (62.6 Mpc) and crowding. We find radii for seven SCCs, with only two of them
being bound objects.

Our results are similar to a ground-based study of the entire body of NGC 6872,
		which found young clusters (< 10 Myr) and objects around $10^6$ \msol\ in both of the tidal tails of the system \citep{bastian_05_2}.
		They also did not find any objects older than 145 Myr in the tails; these resided in the interior of the galaxy. Although we have found
		SCCs older than this in our tidal tail, the majority are less than this age: 18 out of the 158 SCCs are older than 145 Myr.

\subsection{NGC 3256E/W}
NGC 3256 has an IR emission of L$_\textrm{IR}$ = 3.3 x 10$^{11} L_{\odot}$ \citep{lipari_04}, classifying it as a Luminous Infrared Galaxy (LIRG).
This is a starbursting galaxy with two prominent tidal tails.

The central 5 kpc of the system shows three compact knots, one at optical wavelengths and the other two in IR, as well as an asymmetric
spiral arm associated with each individual source, suggesting this system is the result of three galaxies merging together. The likely
scenario involves two disk galaxies merging first, followed by a smaller, satellite galaxy merging \citep{lipari_04,lipari_00}.

The interior shows prodigious amounts of star formation, powering a galactic wind and the system's large IR luminosity \citep{lipari_00}.
This is visualized in bright H\textsc{ii} regions \citep{english_03,lipari_00} and numerous young, blue clusters \citep{mulia_16,goddard_10,zepf_99}. Broadband analysis of such objects shows the age distributions have an overdensity of objects
with ages less than 10 Myr \citep{mulia_16,goddard_10}. Spectroscopy of several of these internal clusters also shows they
have a high metallicity, at Z $\sim$ 1.5 Z$_\odot$ \citep{trancho_07_2}. Similarly, three clusters in the Western tidal tail studied
spectroscopically show similar elevated metallicity levels of Z $\sim$ 1.5 Z$_\odot$ \citep{trancho_07_1}.

SCCs for both NGC 3256W and NGC 3256E do not show the large numbers of young objects seen in the nucleus. There are
only four in the Eastern tail and one in the Western tail that fall below an age of 10 Myr. Populations of SCCs in both tails
have ages comparable to the interaction age of 400 Myr, suggesting they formed in the interaction. \cite{rodruck_16} find similar results
for ages of the diffuse light as well. This is also seen in \cite{mulia_15}, who looked at clusters in NGC 3256E, 
		at the base of the tail, near the nucleus. The ages of these objects were weakly clustered around 250 Myr.

A KS test between the SCCs for the two tails for their \U\ - \B, \B\ - $V$, and $V$ - $I$ colours gives \textit{p}-values of 0.017, 2.2 $\times 10^{-4}$, and 8.3 $\times 10^{-3}$,
respectively. A KS test between their ages and masses gives \textit{p}-values of 0.05 and 2.0 $\times 10^{-4}$, respectively. The KS tests
show differences between the masses of the clusters, though not the ages. NGC 3256E appears to be currently forming clusters,
as also seen in \cite{rodruck_16}.

KS tests for in-tail and out-of-tail SCCs give similar results for both tails. For NGC 3256E, we find $p$-values of 0.92, 0.17, and 0.23 for
\U\ - \B, \B\ - $V$, and $V$ - $I$, respectively. For NGC 3256W, we find $p$-values of 0.97, 0.33, and 0.18 for \U\ - \B, \B\ - $V$, and $V$ - $I$, respectively.
The cluster excess for NGC 3256E is 0.034 $\pm 0.014$ kpc$^{-2}$, and 0.103 $\pm 0.025$ kpc$^{-2}$ for NGC 3256W. The excess
in NGC 3256W is above 3$\sigma$.

SCCs in NGC 3256 show a median radius of 8.5 pc, larger than the typical radii of clusters in extragalactic observations, which generally peak around
3 -- 5 pc \citep{ennis_19,ryon_17,chandar_16,bastian_12,jordan_07}. However, these other studies still see objects with large radii, and such large objects
are expected to form in galaxy mergers \citep{renaud_15}.
All sources for which we have a measured radius appear to be bound, from Figure \ref{fig:radii_pi}. Of particular note is a massive
object located near the tip of the tail, with age $10^{8.3}$ yr, mass $10^{6.3}$ \msol, and a radius of 10.7 pc. This is our fourth largest
SCC in NGC 3256W, and is 50 kpc from the centre of the merging system. This object could evolve to become similar to a present day globular cluster. Clusters
formed in tidal tails can migrate outwards and become isolated globular clusters \citep{matsui_19}.


	NGC 1487 stands out from the rest of the mergers in our sample: it is three times closer than
our next closest galaxy, there is only one SCC between the two tidal tails, and it has been classified
as both a merger between two disk galaxies \citep{aguero_97} and a merger between dwarf galaxies
\citep{buzzo_21,bergvall_03}. Despite the lack
of SCCs, visual examination of the tails shows an abundance of objects within the debris. This suggests
the debris host faint, low-mass objects which belong to the merging system, but are not luminous, high-mass 
clusters with $M_V < -8.6$. The 
absence of high-mass objects may arise if this is indeed a merger between dwarf galaxies.	
Massive star clusters and high SFRs require high gas pressure \citep{maji_17,zubovas_14,blitz_06},
and these required pressures may not be produced in a dwarf galaxy merger. \cite{lahen_19} were
able to simulate a merger between two equal sized dwarf galaxies, which produced clusters with masses
$\ge 10^5$ \msol. Pressures in these clusters was found to be $\sim 10^7$ $k$(K cm$^{-3}$)$^{-1}$, smaller
than the $10^8 - 10^{12}$ $k$(K cm$^{-3}$)$^{-1}$ values seen in simulations of major mergers \citep{maji_17}.

\subsection{NGC 1487E/W} \label{sec:A1487}

NGC 1487 is our nearest merging system in our sample, at 10.8 Mpc. This is a peculiar system in that it 
has been called a disk merger \citep{aguero_97} and a dwarf merger \citep{buzzo_21,bergvall_03}.
\cite{aguero_97} classify it as a merger between disk galaxies due to emission line strength, which gives a SFR higher than seen in isolated galaxies.
\cite{lee_05} give an age for the merger at 500 Myr, as \textit{BI}-band photometry of star clusters in the nucleus suggests they are 500 Myr. This age estimate is supported by integral field spectroscopy \citep{buzzo_21}.
		However, these estimates are based on the assumption that the age of the interaction is correlated to the 
		number density peak in cluster age, implying that they formed due to the interaction. It is possible that the present tidal
		tails could have formed in a secondary encounter between the galaxies, and thus have a younger age than the 
		quoted value.

Evidence of a dwarf merger is seen in its low mass ($4.6 \pm 1.2 \times 10^9$ \msol) and low mass of nuclear concentrations ($\sim 1.5 - 0.2$ $\times$ 10$^9$ \msol),
similar velocity fields between concentrations, and low SFRs \citep{buzzo_21}.

Tidal debris extends to the East and West. Most interestingly, it is the only galaxy in our sample not to contain high-mass clusters. As discussed in Section \ref{sec:1487}, NGC 1487 may be a dwarf
galaxy merger, and not able to produce the necessary pressures to form massive clusters, as
seen in major mergers.
We find only one SCC in NGC 1487W, and none in NGC 1487E. The rest of the detected objects fall below our magnitude cut of $M_V < -8.5$.
This does not mean that it does not contain star clusters, but rather it does not contain high-mass clusters. Despite the lack
of SCCs, both regions of debris show clear excesses of detected objects, as compared to their respective out-of-tail regions.
The cluster excesses for in-tail vs out-of-tail sources is 4.2 $\pm$ 0.6 for NGC 1487W, and 
3.3 $\pm$ 0.6 for NGC 1487E, which are statistically significant for each tail. 

The mass function for objects in the debris shows a similar slope as our SCCs (Figure \ref{fig:MF_1487}), though with smaller
radii (Figure \ref{fig:1487_r}). The implication is that we are seeing the low-mass regime of cluster formation in NGC 1487. We
are able to see these objects despite their faintness as NGC 1487 is relatively nearby (10.8 Mpc).
Three clusters in the central region observed by \cite{mengel_08}
show similar properties to what we see in the tidal tails, with radii between 1 -- 3 pc and masses betwen 7 -- 15 $\times 10^4$ \msol.
\HI\ observations show high densities, at 2.55 $\pm 0.51 \times 10^7$ \msol\ kpc$^{-2}$, but \cite{mullan_11} suggest this
may be an inclination effect. 

The ages of objects found in the tails closely resemble previous studies of star formation in NGC 1487.
		\cite{lee_05} found a bimodal distribution in the colours of star clusters, which correspond to ages of 15 Myr and 500 Myr. \cite{buzzo_21} find
		the star formation history of the system to have peaks at $\sim 10$ and $\sim 300$ Myr. Our data in the tails show objects with ages $< 10$ Myr,
		as well as a clustering of objects at 500 Myr, matching the star formation history of the tails to the interior.





	\bsp	
	\label{lastpage}
\end{document}